\documentclass[nofootinbib,aps,11pt]{revtex4}
\usepackage{graphicx}
\usepackage{cancel}
\usepackage{amssymb}
\usepackage{textcomp}
\usepackage{amsmath}
\usepackage{bm}
\usepackage{times}
\usepackage{epsfig}
\usepackage{color}
\usepackage{graphics}
\def\beq{\begin{equation}}
\def\eeq{\end{equation}}

\def\ev{\,{\rm eV}}
\def\be{\begin{equation}}
\def\ee{\end{equation}}
\def\bea{\begin{eqnarray}}
\def\eea{\end{eqnarray}}
\def\zbl{Z_{B-L}}
\def\dl{\Delta L}
\def\md{m^{}_D}
\def\4j{\rm 4~jets}
\begin{document}
\begin{flushright}
MADPH-09-1534
\end{flushright}
\title{\large TESTABILITY OF TYPE I SEESAW AT THE CERN LHC: \\
REVEALING THE EXISTENCE OF THE $B-L$ SYMMETRY}
\bigskip
\author{Pavel Fileviez P{\'e}rez$^1$\footnote{Electronic address: fileviez@physics.wisc.edu}}
\author{Tao Han$^{1,2}$\footnote{Electronic address: than@hep.wisc.edu}}
\author{Tong Li$^2$ \footnote{Electronic address: nklitong@hotmail.com, communication author}}
\address{
$^{1}$Department of Physics, University of Wisconsin, Madison, WI 53706, USA\\
$^{2}$Center for High Energy Physics, Peking University, Beijing
100871, P.R.~China}
\date{\today}

\begin{abstract}
We study the possibility to test the Type I seesaw mechanism for
neutrino masses at the CERN Large Hadron Collider. The inclusion of
three generations of right-handed neutrinos ($N_i$) provides an
attractive option of gauging the $B-L$ accidental symmetry in the
Standard Model (as well as an extended symmetry $X=Y-5(B-L)/4$). The
production mechanisms for the right-handed neutrinos through the
$Z'$ gauge boson in the $U(1)_{B-L}$ and $U(1)_X$ extensions of the
Standard Model are studied. We discuss the flavor combinations of
the charged leptons from the decays of $N_i$ in the $\Delta L=2$
channels.
We find that the clean channels with dilepton plus jets and possible
secondary vertices of the $N$ decay could provide conclusive signals
at the LHC in connection with the hierarchical pattern of the light
neutrino masses and mixing properties within the Type I seesaw
mechanism.
\end{abstract}
\maketitle
\section{Introduction}
The small but non-zero neutrino masses lead  to a deep conjecture:
Majorana nature of the neutrino masses may hold the key for a
fundamentally different mass generation mechanism, although Dirac
masses can be generated via the Higgs mechanism by introducing
right-handed neutrinos with arbitrarily small Yukawa couplings.
There are three simple scenarios where one can generate Majorana
masses of the neutrinos with renormalizable operators at tree level,
and in agreement with the observations, the Type I~\cite{TypeI},
Type II~\cite{TypeII}, and Type III~\cite{TypeIII} seesaw
mechanisms. See also Refs.~\cite{Zee} and \cite{wise} for the
simplest neutrino mass generation mechanisms using radiative
corrections.

Perhaps the simplest and best-studied mechanism for neutrino masses
is the Type I seesaw,  where one introduces at least two
right-handed neutrinos $(N)$. Adding in the corresponding large
Majorana mass terms ($M$), one results in at least two light
Majorana neutrinos with masses given as $m_D^2/M$. It is important
to mention that the inclusion of three right-handed neutrinos also
provides an anomaly-free formulation for a gauged U(1)$_{B-L}$
\cite{B-L}.

The non-ambiguous test of the Majorana nature of the neutrinos, and
thus a possible test of the seesaw mechanism, will be the
observation of the lepton number violation processes. The
neutrinoless double beta decay is also a crucial test and one of the
most sensitive probes. Since the CERN Large Hadron Collider (LHC) is
going to lead us to a new energy frontier, searching for the heavy
Majorana neutrinos at the LHC appears to be very
appealing~\cite{senjanovic,taotypei}.
However, due to the rather small mixing between the heavy neutrinos
and the Standard Model (SM) leptons in a minimal Type I scheme,
typically of the order $|V_{\ell N}|^2\sim m_\nu/M_N$, the predicted
effects of lepton number violation are unlikely to be observable. On
the other hand, if there are other particles beyond the SM that can
mediate new interactions between them, the effects may be
significantly enhanced.
For instance, with the new gauge interaction U(1)$_{B-L}$, the gauge
boson $\zbl$ can be produced copiously at the LHC via its gauge
interactions with the quarks. Its subsequent decay to a pair of
heavy Majorana neutrinos may lead to a large sample of events
without involving the small mixing angle suppression of
$N$~\cite{Huitu,Hung:2006ap}.  The $\dl =2$ signals will directly test its
Majorana nature; and the lepton flavor combination could probe the
properties of the light neutrino mass spectrum and mixing pattern.

In this paper, we investigate the possibility to test the Type I
seesaw mechanism at the LHC in the context of two simple extensions
of the Standard Model where one has an extra Abelian gauge symmetry.
We focus our attention on scenarios with a $U(1)_{B-L}$ or $U(1)_X\ (X=Y-5(B-L)/4)$,
where $B$, $L$ and $Y$ stand for Baryon number, Lepton number and weak hypercharge,
respectively.
In order to cancel the anomalies, one just need to introduce three right-handed
neutrinos, which are the source for the Majorana masses.
In both scenarios one has a new neutral gauge boson,
$Z'$, which couples to the right-handed neutrinos. Then, one can
expect large number events for the lepton number violating events
due to the production and decays of the TeV Majorana neutrinos. The
predictions of the heavy neutrino decays in each neutrino spectrum,
Normal Hierarchy (NH), Inverted Hierarchy (IH) or Quasi-Degenerate
(QD), are investigated in great detail. We find encouraging results for
the LHC signatures to learn about the light neutrino properties.

This work is organized as follows: In Section II we discuss the
constraints on the mass and mixing parameters in the Type I seesaw
mechanism from the current neutrino oscillation data. The
predictions for the decays of the heavy neutrinos in the different
neutrino spectra are presented in Section III. In Section IV we
discuss the possibility to test Type I seesaw at the LHC through the
same-sign dilepton channels. We summarize our findings in Section V.
The mixing between light and heavy neutrinos are discussed in
Appendix A. We provide the explicit expressions for these mixings in Appendix B.
The minimal extensions of the Standard Model to $U(1)_{B-L}$ and $U(1)_{X}$
are discussed in Appendix C.

\section{Type I Seesaw Mechanism and parameter constraints}
In the case of the Type I seesaw mechanism for neutrino masses one introduces at least two
SM singlets, right-handed neutrinos, $\nu_R \sim (1,1,0)$, in order to generate two non-zero
neutrino masses. In this case the relevant Yukawa interaction and the Majorana mass term are given by
\begin{eqnarray}
    - {\cal L}_{\nu}^I  &  =  & Y_\nu^D \ \bar{l}_L \ \tilde{H} \ \nu_R
\ + \ \frac{M_N}{2} \ \nu_R^T \ C \ \nu_R + \ \rm{h.c.} .
\end{eqnarray}
Here $\tilde{H}=i\sigma_2 H^*$ and the lepton number is broken
in two units  due to the presence of both terms. Now, integrating
out the right-handed neutrinos one finds that the mass
matrix for the light neutrinos is given by
\begin{equation}
M_\nu = m_D \ M^{-1}_N \ m_D^T,
\label{numass}
\end{equation}
where $m_D=Y_\nu^D v_0/\sqrt{2}$ is the Dirac mass term
and $v_0$ is the Higgs vacuum expectation value.
Therefore, in this framework one could understand the smallness of
neutrino masses, since the mass scale $M_N$ in the
above equation could be large, $M_N \gg Y_\nu^D v_0$,
This is the so-called canonical Type I seesaw mechanism~\cite{TypeI}.
 The mass matrix for neutrinos is diagonalized by unitary rotations as detailed
 in Appendix A. The three light neutrino masses can be expressed in the following way
\begin{equation}
m = V_{PMNS}^\dagger \ M_\nu \ V_{PMNS}^*,
\end{equation}
where $m=diag (m_1, m_2, m_3)$ and $V_{PMNS}$ can be taken as  the
leptonic mixing matrix for the three generation of light
neutrinos~\cite{PMNS} without the loss of generality.\footnote{The
$3\times 3$ rotational matrix is not exactly unitary when there are
extra Majorana neutrinos, but it is a good approximation to equal it
to the traditional $V_{PMNS}$, see the formalism in the appendix.}
Working in the basis where the heavy neutrino mass matrix is
diagonal and using the Casas-Ibarra
parametrization~\cite{Casas:2001sr} one can write $m_D$ satisfying
Eq.~(\ref{numass}) as
\begin{equation}
m_{D}= V_{PMNS} \ m^{1/2} \ \Omega \ M^{1/2},
\label{Dirac}
\end{equation}
where  $M=diag(M_1, M_2, M_3)$ for heavy neutrino masses,
and $\Omega$ is a complex matrix which satisfies the orthogonality
condition $\Omega^T \Omega = 1$.  It is shown in
Appendix A that using the seesaw formula and the relation between
the leptonic mixing one can find a formal solution for the mixing
between the SM charged leptons ($\ell=e, \mu, \tau$) and heavy neutrinos
($N=1,2,3$):
\begin{eqnarray}
V_{\ell N}= \ V_{PMNS} \ m^{1/2} \ \Omega \ M^{-1/2}.
\label{mixing1}
\end{eqnarray}
Therefore, for a given form of $\Omega$, one can establish the
connection between the heavy neutrino decays and the properties of
the light neutrinos.
The impact of the existence of the $\Omega$ matrix on the decays of
heavy neutrinos has not been studied before in collider
phenomenology. Unfortunately, since the explicit form of this matrix
is unknown one cannot predict the decay pattern of the heavy
neutrinos with respect to the spectrum for light neutrinos.
We will present a few well-motivated typical cases where one can
hope to see the connection in each spectrum for light neutrinos.
It is important, however,  to realize that an
underlying theory would pick only one specific form
of $\Omega$. This (yet unknown) form
would have definite prediction for the $N$ decay patterns, through which
the underlying theory could be revealed.

\subsection{Constraints on the Physical Parameters}
\subsubsection{\bf Neutrino Masses and Mixings}
In order to understand the constraints coming from neutrino physics
let us discuss the relation between the neutrino masses and mixing.
The leptonic mixing matrix is given by
\begin{equation}
V_{PMNS}= \left(
\begin{array}{lll}
 c_{12} c_{13} & c_{13} s_{12} & e^{-\text{i$\delta $}} s_{13}
   \\
 -c_{12} s_{13} s_{23} e^{\text{i$\delta $}}-c_{23} s_{12} &
   c_{12} c_{23}-e^{\text{i$\delta $}} s_{12} s_{13} s_{23} &
   c_{13} s_{23} \\
 s_{12} s_{23}-e^{\text{i$\delta $}} c_{12} c_{23} s_{13} &
   -c_{23} s_{12} s_{13} e^{\text{i$\delta $}}-c_{12} s_{23} &
   c_{13} c_{23}
\end{array}
\right)\times \text{diag} (e^{i \Phi_1/2}, 1, e^{i \Phi_2/2})
\end{equation}
where $s_{ij}=\sin{\theta_{ij}}$, $c_{ij}=\cos{\theta_{ij}}$, $0 \le
\theta_{ij} \le \pi/2$ and $0 \le \delta \le 2\pi$. The phase
$\delta$ is the Dirac CP phase, and $\Phi_i$ are the Majorana
phases. The experimental constraints on the neutrino masses and
mixing parameters, at $2\sigma$ level~\cite{Schwetz}, are
\bea
7.25 \times 10^{-5} \ev^2 \  < & \Delta m_{21}^2 & < \  8.11 \times 10^{-5} \ev^2, \\
2.18 \times 10^{-3} \ev^2 \  < & |\Delta m_{31}^2| & < \  2.64 \times 10^{-3} \ev^2, \\
                   0.27 \  < & \sin^2{\theta_{12}} & < \  0.35, \\
                   0.39 \  < & \sin^2{\theta_{23}} & <\  0.63, \\
                          & \sin^2{\theta_{13}} & <\  0.040,
\eea
and $\sum_{i} m_{i} < \ 1.2 \ \ev$. For a complete discussion of
these constraints see reference~\cite{review}. Following the
convention, we denote the case $\Delta m_{31}^2 > 0$ as the normal
hierarchy (NH); $\Delta m_{31}^2 < 0$  the inverted hierarchy (IH),
and the quasi-degenerate (QD) spectrum where the lightest neutrino mass
is larger than $5 \times 10^{-2}$ eV. Using the above experimental constraints,
one can expect to explore the
allowed values for the $V_{\ell N}$ couplings and the heavy masses.
From Eq.~(\ref{mixing1}),  we can obtain the general expressions of
$\sum_{N}(V_{\ell N}^\ast)^2$ that are collected in Appendix B.

\subsubsection{\bf Case I: Degenerate Heavy Neutrinos}
We firstly study the simplest case where the three heavy neutrinos
are degenerate. This is a highly motivated scenario since it is
strongly favored to generate successful
resonant lepto-genesis~\cite{resonant,Steve} at the low scale.
Using Eq.~(\ref{VLN}) and assuming degenerate heavy neutrinos we
obtain the relation
\begin{eqnarray}
M \sum_{N=1,2,3} \ (V_{\ell N}^\ast)^2&=&(V^\ast_{PMNS}\ m \
V^\dagger_{PMNS})_{\ell\ell}\equiv (M_\nu)_{\ell\ell} \ , \  \
(\ell=e,\mu,\tau). \label{degener}
\end{eqnarray}
We see that one can obtain simple relations for the heavy neutrino
mixings and masses in terms of the light neutrino mass matrix
independent of the unknown matrix $\Omega$, which in turn is given
by the parameters from the neutrino oscillation data. One can thus
predict the decays of the heavy neutrinos in each spectrum for light
neutrinos. Note that in this degenerate scenario,
we are unable to convert the constraints
of Eq.~(\ref{degener}) to predict $\sum_N |V_{\ell N}|^2$ in general.
We can predict the decays of heavy neutrinos in terms of the other
oscillation parameters only when all phases vanish since in this case the modulo
square of the mixings (which govern the decay rate) are equal to the
square of mixings (the left-handed side of Eq.~(\ref{degener})).

\begin{figure}[tb]
\begin{center}
\begin{tabular}{cc}
\includegraphics[scale=1,width=8cm]{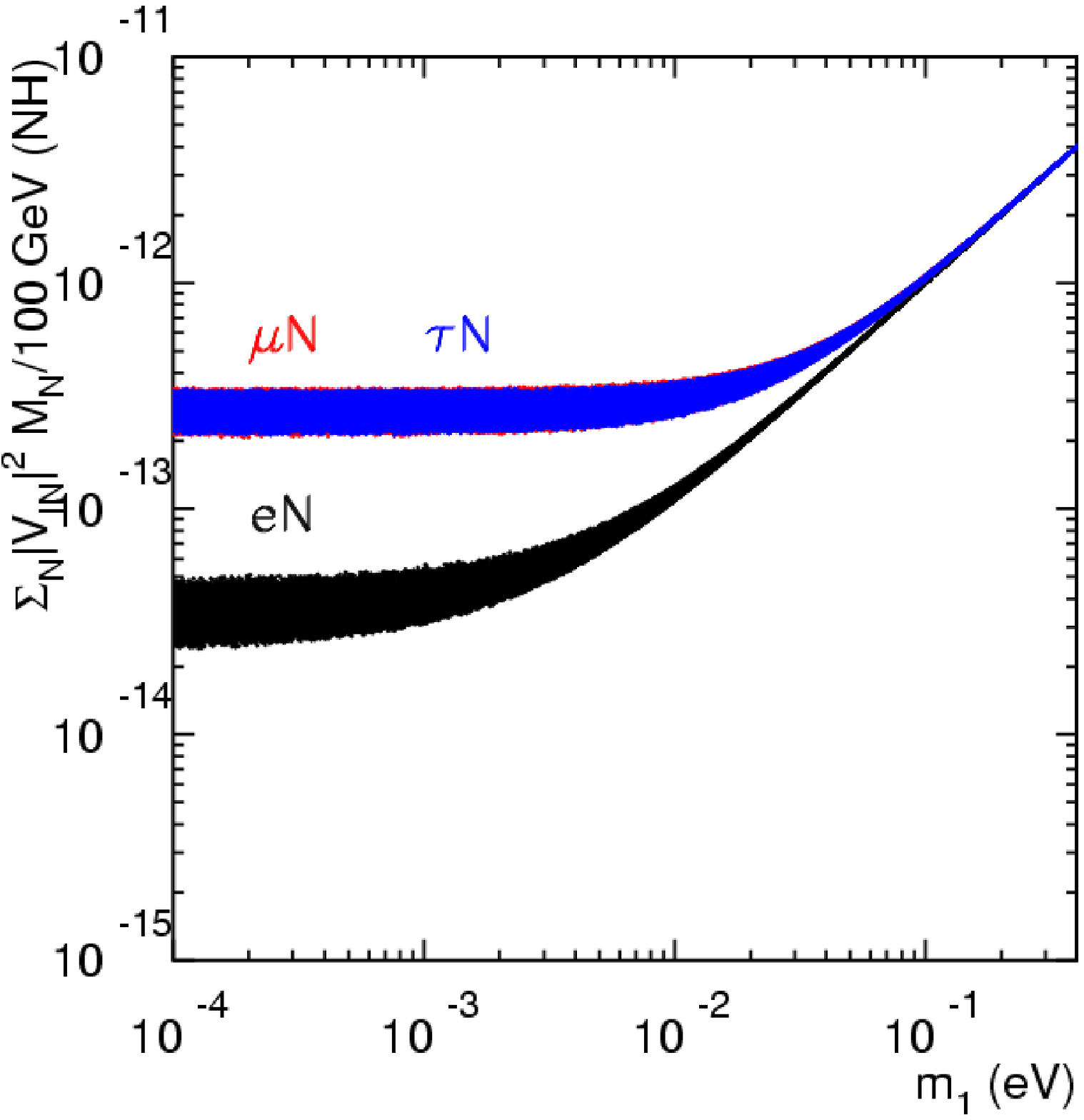}
\includegraphics[scale=1,width=8cm]{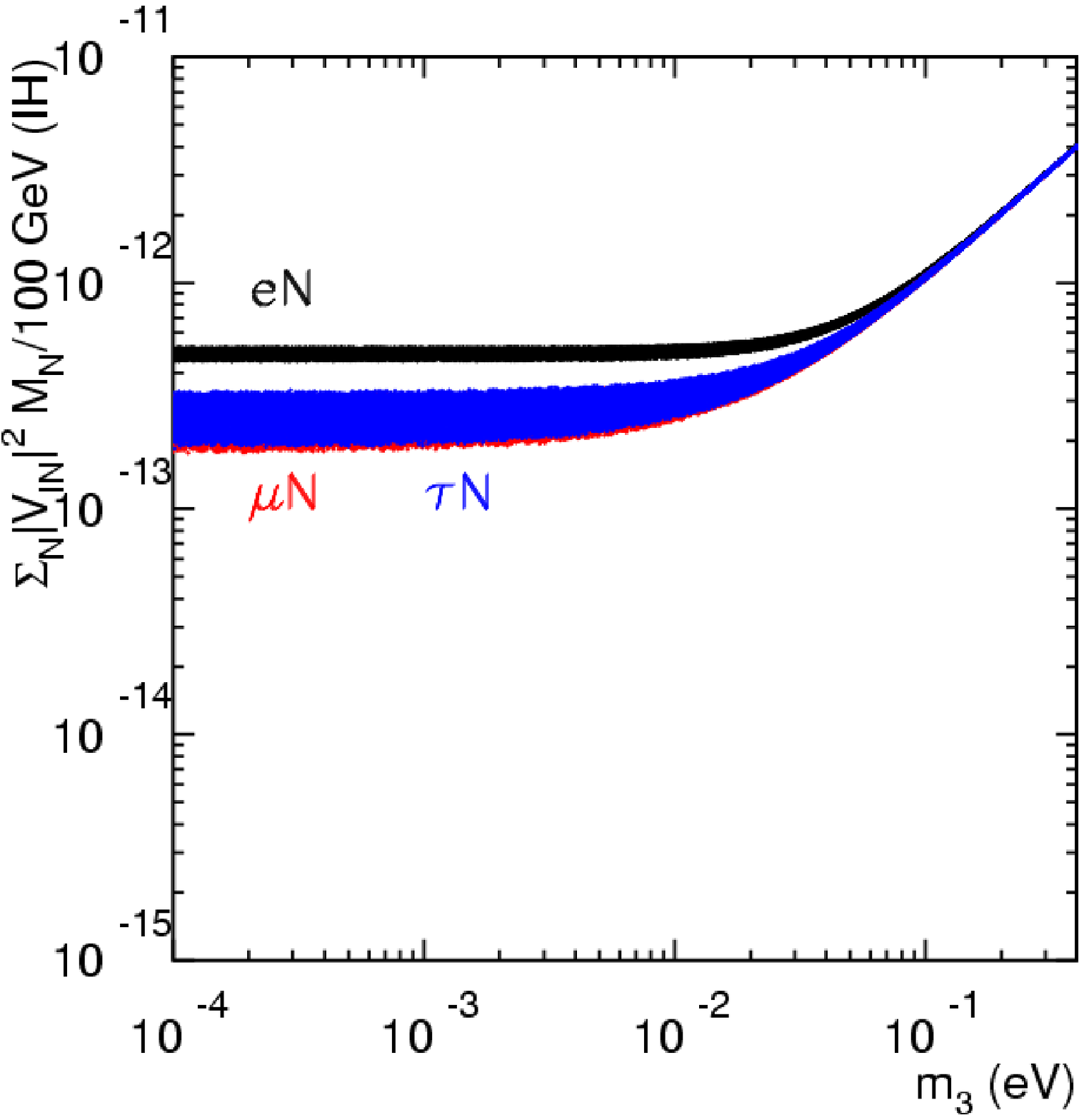}
\end{tabular}
\end{center}
\caption{$\sum_N|V_{\ell N}|^2 M_N/100~{\rm GeV}$ versus the
lightest neutrino mass for NH (left) and IH (right) in Case I
(degenerate $N$), assuming vanishing Majorana phases.} \label{VLlN}
\end{figure}

In Ref.~\cite{ourtype2}, we have shown that using the experimental constraints on the neutrino mass parameters
the elements of the neutrino mass matrix has the following properties:
\begin{eqnarray}
&&M_\nu^{ee}\ll M_\nu^{\mu\mu}, M_\nu^{\tau\tau} \ \ \ {\rm for \ \ NH},\nonumber \\
&&M_\nu^{ee}> M_\nu^{\mu\mu}, M_\nu^{\tau\tau} \ \ \ {\rm for \ \
IH},\\
&&M_\nu^{ee}\approx M_\nu^{\mu\mu}\approx M_\nu^{\tau\tau} \ \  \
{\rm for \ \ QD}.
\label{hiera}
\end{eqnarray}
Following the same approach, we plot the allowed values for the
normalized couplings of each lepton flavor in this scenario in
Fig.~\ref{VLlN}, as a function of the lightest neutrino mass in each
spectrum, the normal hierarchy (left panel) and the inverted
hierarchy (right panel), assuming vanishing Majorana phases.  We see
two distinctive regions in terms of the lightest neutrino mass as
expected.
In the case $m_{1(3)}  < 5\times 10^{-2}$ eV, we see the characteristic features
\begin{eqnarray}
&& \sum_{N} \ |V_{eN}|^2 \ll \sum_{N} \ |V_{\mu N}|^2, \sum_{N} \ |V_{\tau N}|^2
\ \ \ {\rm for \ \ NH},\nonumber \\
&& \sum_{N} \ |V_{eN}|^2 \ >  \ \sum_{N} \ |V_{\mu N}|^2, \sum_{N} \ |V_{\tau N}|^2
\ \ \ {\rm for \ \ IH}.
\nonumber
\end{eqnarray}
On the other hand, for $m_{1(3)}  > 5\times 10^{-2}$ eV, the light neutrino masses enter the
QD spectrum that leads to
\begin{equation}
\sum_{N} \ |V_{eN}|^2\approx \sum_{N} \ |V_{\mu N}|^2 \approx \sum_{N} \ |V_{\tau N}|^2.
\nonumber
\end{equation}
Under this mass degenerate assumption, the mixing between the heavy
neutrinos and the SM charged leptons simply reflect the features of
the light neutrino mass matrix in the flavor basis, as seen in
Eq.~(\ref{degener}). This is an important model-prediction. It is
important to emphasize that the results shown in Fig.~\ref{VLlN} may
be used to learn about the neutrino spectrum.

%
\subsubsection{\bf Case II: Non-degenerate Heavy Neutrinos}
If we relax the assumption that heavy Majorana neutrinos are nearly degenerate in mass,
then the complication due to the unknown matrix  $\Omega$ arises. The explicit parameterization
of $\Omega$ is presented in Appendix A, and the general expression for the relations among the
parameters are given in Appendix B.

\begin{figure}[tb]
\begin{center}
\begin{tabular}{cc}
\includegraphics[scale=1,width=8cm]{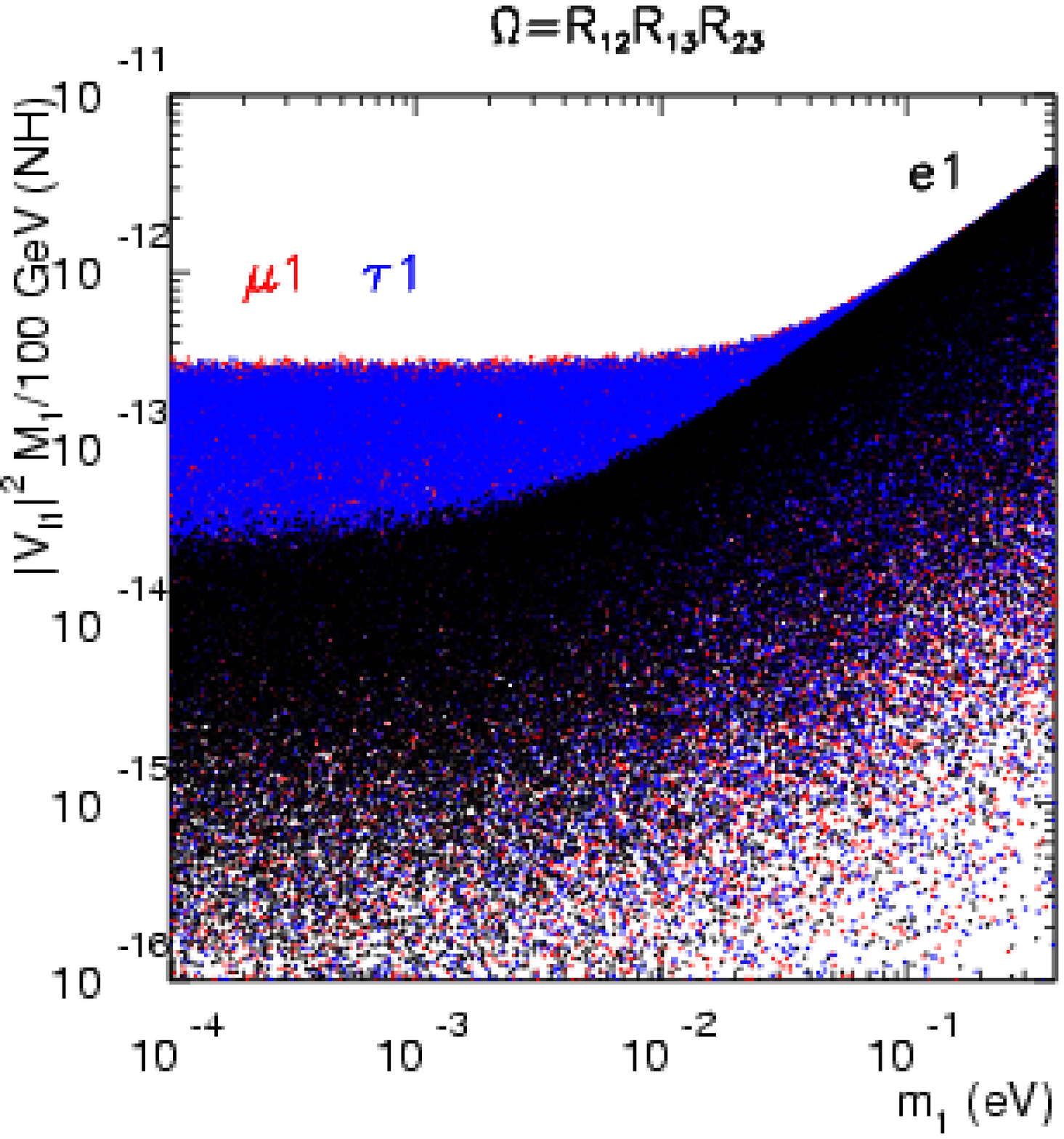}
\includegraphics[scale=1,width=8cm]{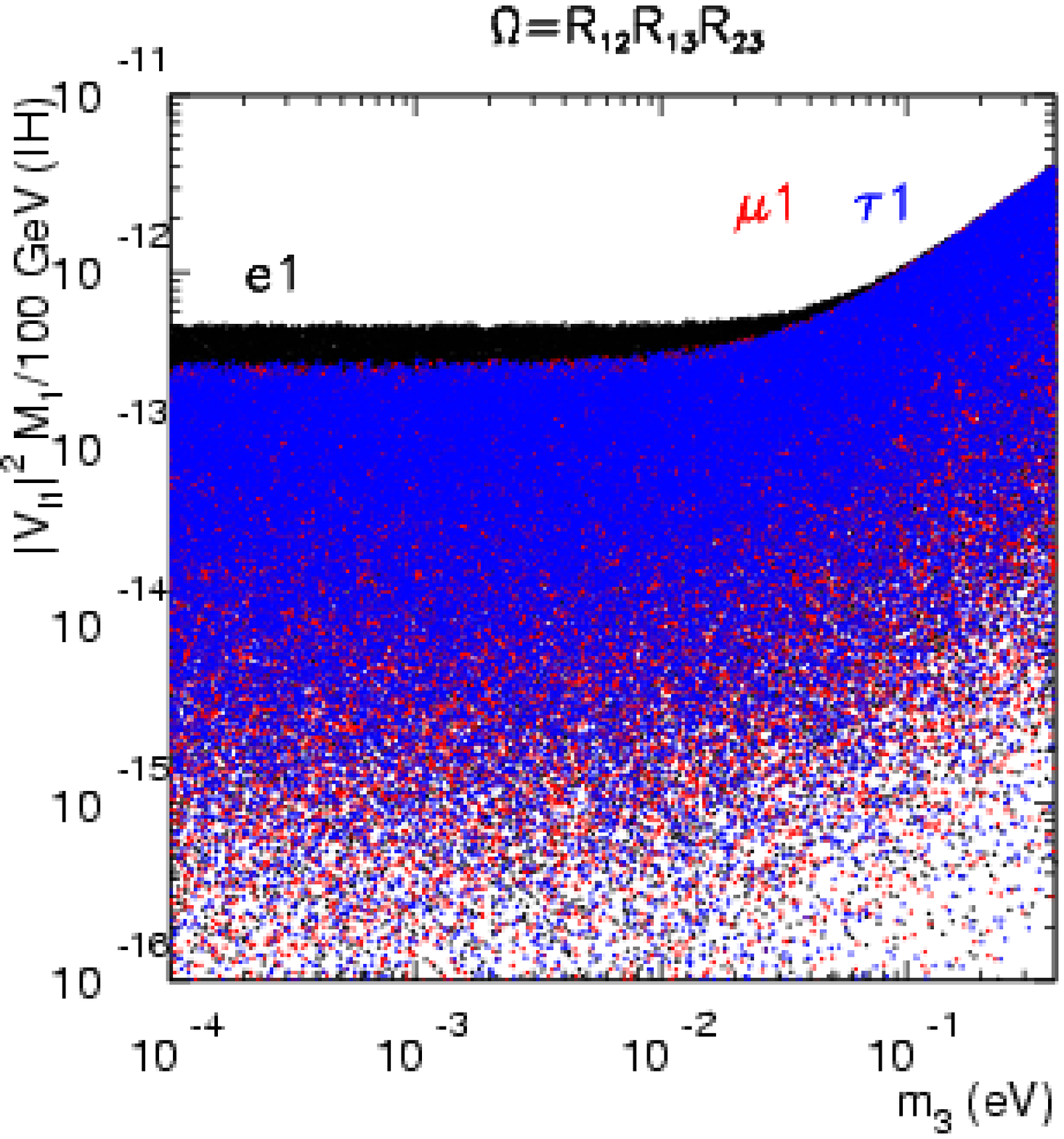}
\end{tabular}
\end{center}
\caption{$|V_{\ell 1}|^2 M_1/100~{\rm GeV}$ versus the lightest
neutrino mass for NH (left) and IH (right) with
$\Omega=R_{12}R_{13}R_{23}$ and random matrix elements
 $-1 \leq w_{ij} \leq 1$,
assuming vanishing Majorana phases.} \label{yl1m1all}
\end{figure}

\begin{figure}[tb]
\begin{center}
\begin{tabular}{cc}
\includegraphics[scale=1,width=12cm]{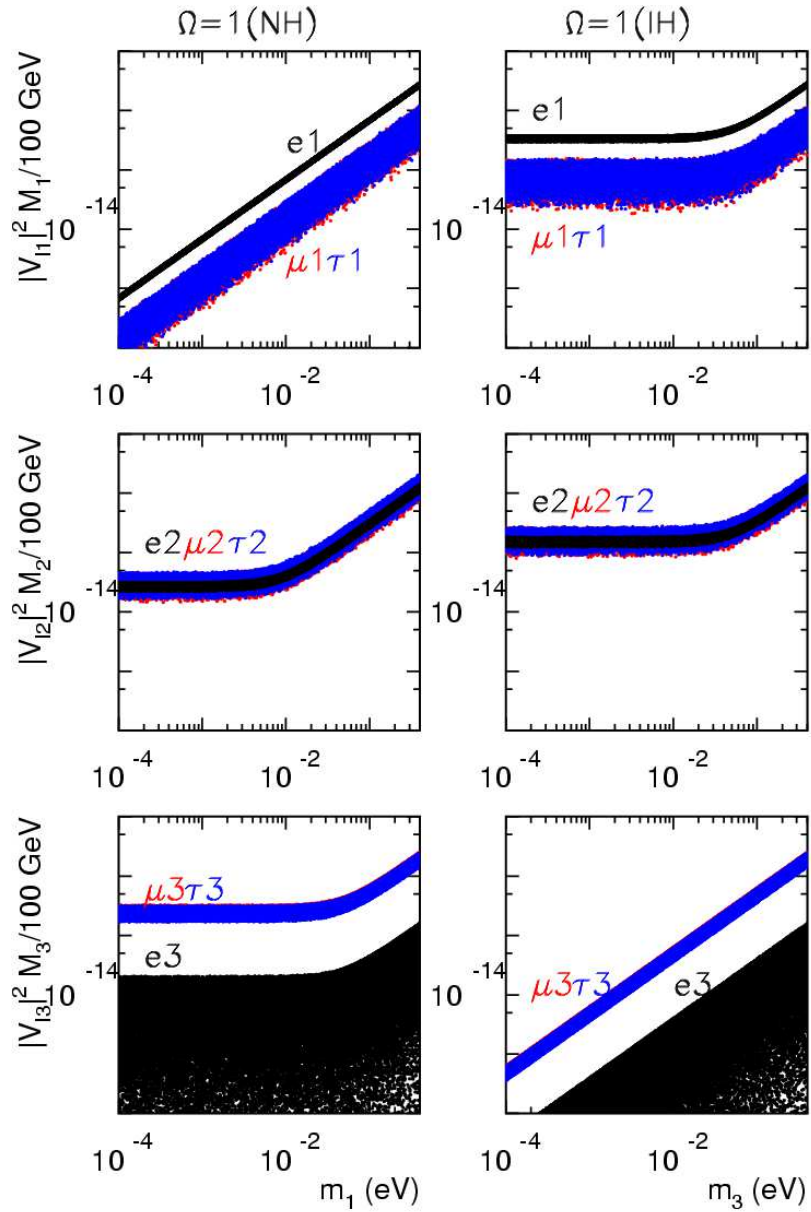}
\end{tabular}
\end{center}
\caption{ $|V_{\ell i}|^2 M_i/100~{\rm GeV}, i=1,2,3$ versus the
lightest neutrino mass for NH (left panels) and IH (right panels) in
Case IIa ($\Omega=I$).} \label{ylimi}
\end{figure}

For the purpose of illustration,
let us take $\Omega$ to be a real matrix. We could gain a general sense for the
mixing parameters by varying the matrix elements of $\Omega$ in the range of
 $-1 \leq w_{ij} \leq 1$.
We show $|V_{\ell1}|^2 M_1/100~{\rm GeV}$ in this case in
Fig.~\ref{yl1m1all}. The predictions of $|V_{\ell2}|^2 M_2/100~{\rm
GeV}$ and $|V_{\ell3}|^2 M_3/100~{\rm GeV}$ are almost the same. As
one can see that qualitative features for both cases of NH and IH
closely resemble that in Fig.~\ref{VLlN}. This is quite encouraging
since the random selection of the model parameters do not seem to
totally wash out the predicted features. To further explore the
model implications, we must choose a specific form of the $\Omega$
matrix, which should correspond to a particular theoretical
incarnation in the right-handed neutrino sector. However, a large
Majorana phase could alter the predictions~\cite{ourtype2} in
general. We will check on this point in the next section.
\centerline
{\bf Case IIa: $\Omega=I$}
In this simple scenario, we easily obtain transparent  relations
for the $N_1$ mixings:
\begin{eqnarray}
|V_{e1}|^2M_1&=&m_1c_{12}^2c_{13}^2\approx m_1c_{12}^2,\\
|V_{\mu 1}|^2M_1&=&m_1|s_{12}c_{23}+c_{12}s_{13}s_{23}e^{i\delta}|^2\approx m_1s_{12}^2c_{23}^2,\\
|V_{\tau
1}|^2M_1&=&m_1|s_{12}s_{23}-c_{12}s_{13}c_{23}e^{i\delta}|^2\approx
m_1s_{12}^2s_{23}^2,
\end{eqnarray}
and therefore $|V_{e1}|^2 \ > \ |V_{\mu 1}|^2, \ |V_{\tau 1}|^2$.
In the case of $N_2$ mixing:
\begin{eqnarray}
|V_{e2}|^2M_2&=&m_2c_{13}^2s_{12}^2\approx m_2s_{12}^2,\\
|V_{\mu 2}|^2M_2&=&m_2|c_{12}c_{23}-s_{12}s_{13}s_{23}e^{i\delta}|^2\approx m_2c_{12}^2c_{23}^2,\\
|V_{\tau
2}|^2M_2&=&m_2|s_{12}s_{13}c_{23}e^{i\delta}+c_{12}s_{23}|^2\approx
m_2 c_{12}^2s_{23}^2,
\end{eqnarray}
and $|V_{e2}|^2\approx |V_{\mu 2}|^2\approx |V_{\tau 2}|^2$.
As for the  $N_3$ mixing,
\begin{eqnarray}
|V_{e3}|^2M_3&=&m_3s_{13}^2\approx 0,\\
|V_{\mu 3}|^2M_3&=&m_3c_{13}^2s_{23}^2\approx m_3s_{23}^2,\\
|V_{\tau 3}|^2M_3&=&m_3c_{13}^2c_{23}^2\approx m_3c_{23}^2.
\end{eqnarray}
and one can see $|V_{\mu 3}|^2,|V_{\tau 3}|^2>|V_{e3}|^2$. These features
are shown in Fig.~\ref{ylimi}.

A few remarks are in order.  First of all, with this choice of a
diagonal matrix $\Omega$, the mixing angle squared $|V_{\ell i}|^2$
for $N_i$ is always proportional to the corresponding light neutrino
mass $m_i$. Consequently, the relative fractions of the mixing to
different lepton flavors are universal for both NH and IH.
Secondly, the Majorana phases do not appear in $|V_{\ell N}|^2$ due
to the special structure of $\Omega$. Thirdly, as seen in
Fig.~\ref{ylimi}, the relative strength of the mixing to different
lepton flavors for each $N_i$ closely follow that for the light
neutrino mass eigenstates. In fact, very much like the light
neutrino mass eigenstate labeling, this should be the defining
feature to label $N_1, N_2$ and $N_3$, if we do not like the less
illuminating ordering $M_1<M_2<M_3$.

\begin{figure}[tb]
\begin{center}
\begin{tabular}{cc}
\includegraphics[scale=1,width=12cm]{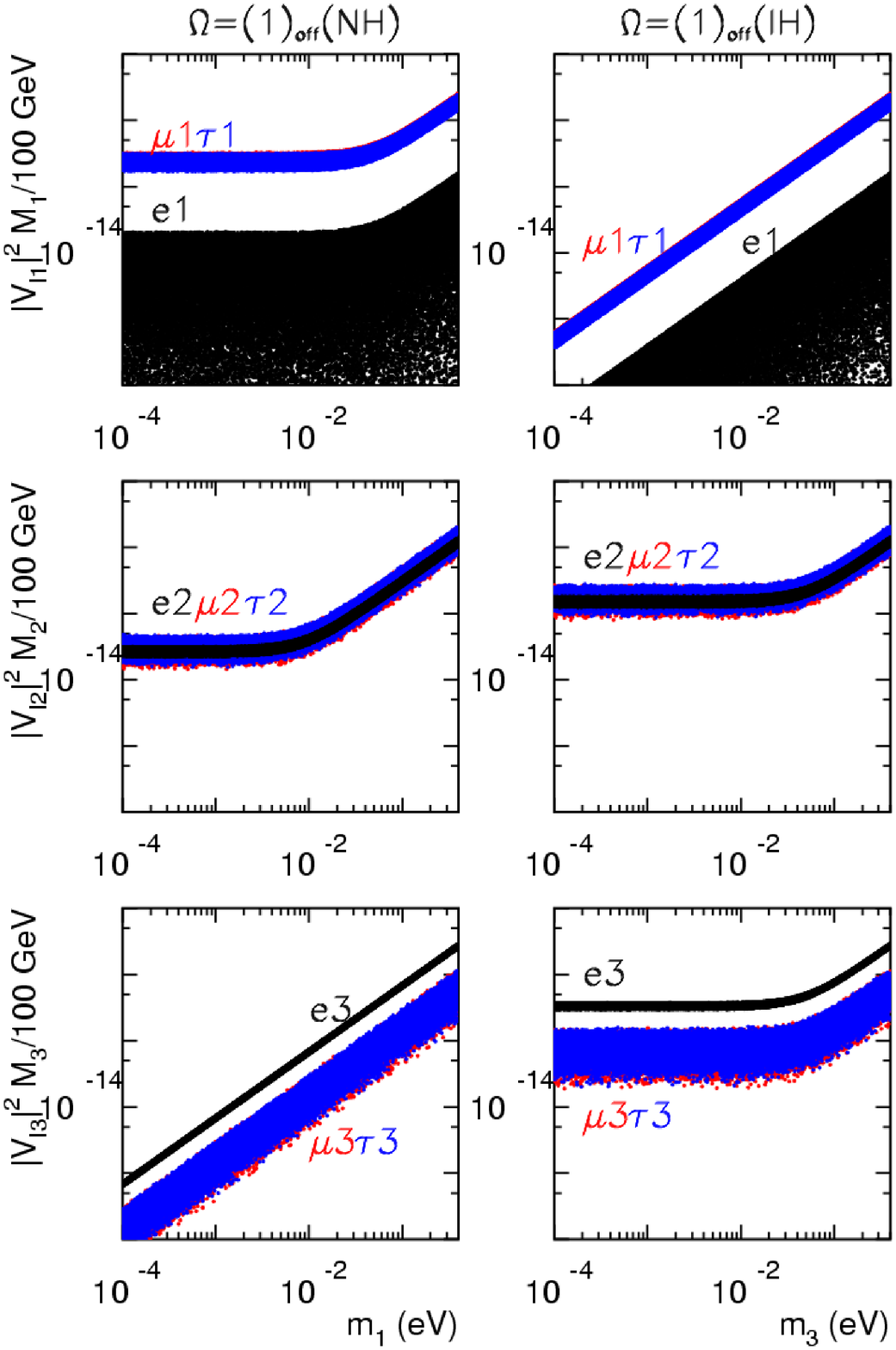}
\end{tabular}
\end{center}
\caption{$|V_{\ell i}|^2 M_i/100~{\rm GeV}, i=1,2,3$ versus the
lightest neutrino mass for NH (left panels) and IH (right panels) in
Case IIb ($\Omega=I_{\rm off}$). } \label{ylimioff}
\end{figure}

\centerline{\bf Case IIb: $\Omega = I_{\rm off}$}
We choose to study yet another simple, but different form of the
matrix, namely, with $\Omega$ as an off-diagonal unity matrix. As
can be shown explicitly and supported by Fig.~\ref{ylimioff},  the
mixing features of $|V_{\ell 1}|^2$ and $|V_{\ell 3}|^2$ switch
places with each other in both NH and IH, while $|V_{\ell 2}|^2$
remain the same as in $\Omega=I$ case. If we recall the convention
for the $N_i$ labeling, this case is indistinguishable from Case
IIa. In this case $|V_{\ell N}|^2$ are also independent of
Majorana phases. This similarity can be generalized to a matrix of
$\Omega$ which has only unity as entries.
We would expect that the real situation could be a well-defined
superposition of the three vertical panels as long as $\Omega$ is real.

\section{Heavy Neutrino Decays and Light Neutrino Spectra}
The leading decay channels for the heavy neutrinos include $N_i \to e^{\pm}_j W^{\mp}$, $N_i \to
\nu_j Z$ and $N_i \to  \nu_j h(H)$. The amplitude for the two first channels are proportional to the mixing
between the leptons and heavy neutrinos given in Eq.~(\ref{mixing1}), while the last one is proportional to
the Dirac-like
Yukawa terms given in Eq.~(\ref{Dirac}).

\subsection{\bf Decay Modes of Heavy Majorana Neutrinos with mass: $M_i \ > \  M_{W} $}
The partial decay widths of the heavy Majorana neutrinos $N_i$ are given by
\begin{eqnarray}
\Gamma^{\ell W_L}&\equiv&
\Gamma(N_i \to \ell^-W_L^+)=
\Gamma(N_i \to \ell^+W_L^-)= {g^2\over 64\pi
M_W^2}|V_{\ell i}|^2M_i^3(1-\mu_{iW})^2,
\\
\Gamma^{\ell W_T}&\equiv&\Gamma(N_i \to \ell^-W_T^+)={g^2\over
32\pi}|V_{\ell i}|^2M_i(1-\mu_{iW})^2,
\\
\Gamma^{\nu_\ell Z_L}&\equiv&\Gamma(N_i\to \nu_\ell Z_L)={g^2\over
64\pi M_W^2}|V_{\ell i}|^2M_i^3(1-\mu_{iZ})^2,
\\
\Gamma^{\nu_\ell Z_T}&\equiv&\Gamma(N_i \to \nu_\ell Z_T)={g^2\over
32\pi c_W^2}|V_{\ell i}|^2M_i(1-\mu_{iZ})^2,
\end{eqnarray}
where $\mu_{ij}=M_j^2/M_i^2$.
If $N_i$ is heavier than the Higgs bosons $h$ and $H$ (see Appendix C for the
properties of the Higgs bosons in the $B-L$ extension of the SM), one has
the additional channels
\begin{eqnarray}
\Gamma^{\nu_\ell h}&\equiv& \Gamma(N_i \to \nu_\ell h)={g^2\over
64\pi M_W^2}|V_{\ell i}|^2M_i^3(1-\mu_{ih})^2\cos^2\theta_0,
\\
\Gamma^{\nu_\ell H}&\equiv& \Gamma(N_i \to \nu_\ell H)={g^2\over
64\pi M_W^2}|V_{\ell i}|^2M_i^3(1-\mu_{iH})^2\sin^2\theta_0 .
\end{eqnarray}
Therefore, the total width for $N_i$ is given by
\begin{eqnarray}
\Gamma_{N_i}&=&\sum_\ell \left(2\Gamma^{\ell W_L}+2\Gamma^{\ell
W_T}+\Gamma^{\nu_\ell Z_L}+\Gamma^{\nu_\ell Z_T}+\Gamma^{\nu_\ell
h}+\Gamma^{\nu_\ell H}\right).
\end{eqnarray}
At a high mass of  $M_N$, the branching ratios of the leading channels go like
\begin{equation}
\Gamma(\ell^-W_L^+) \approx
\Gamma(\ell^+W_L^-) \approx
\Gamma(\nu Z_L) \approx
\Gamma(\nu h+\nu H).
\end{equation}
As discussed above, the lepton-flavor contents of $N$ decays will be
different in each neutrino spectrum. Here, we also study this issue
in great detail for cases I and II. In order to search for the events with
best reconstruction, we will only consider the $N$ decay to charged
leptons plus a $W^\pm$.

\subsubsection{\bf Decays in Case I: Degenerate Heavy Neutrinos}
\begin{figure}[tb]
\begin{center}
\begin{tabular}{cc}
\includegraphics[scale=1,width=8cm]{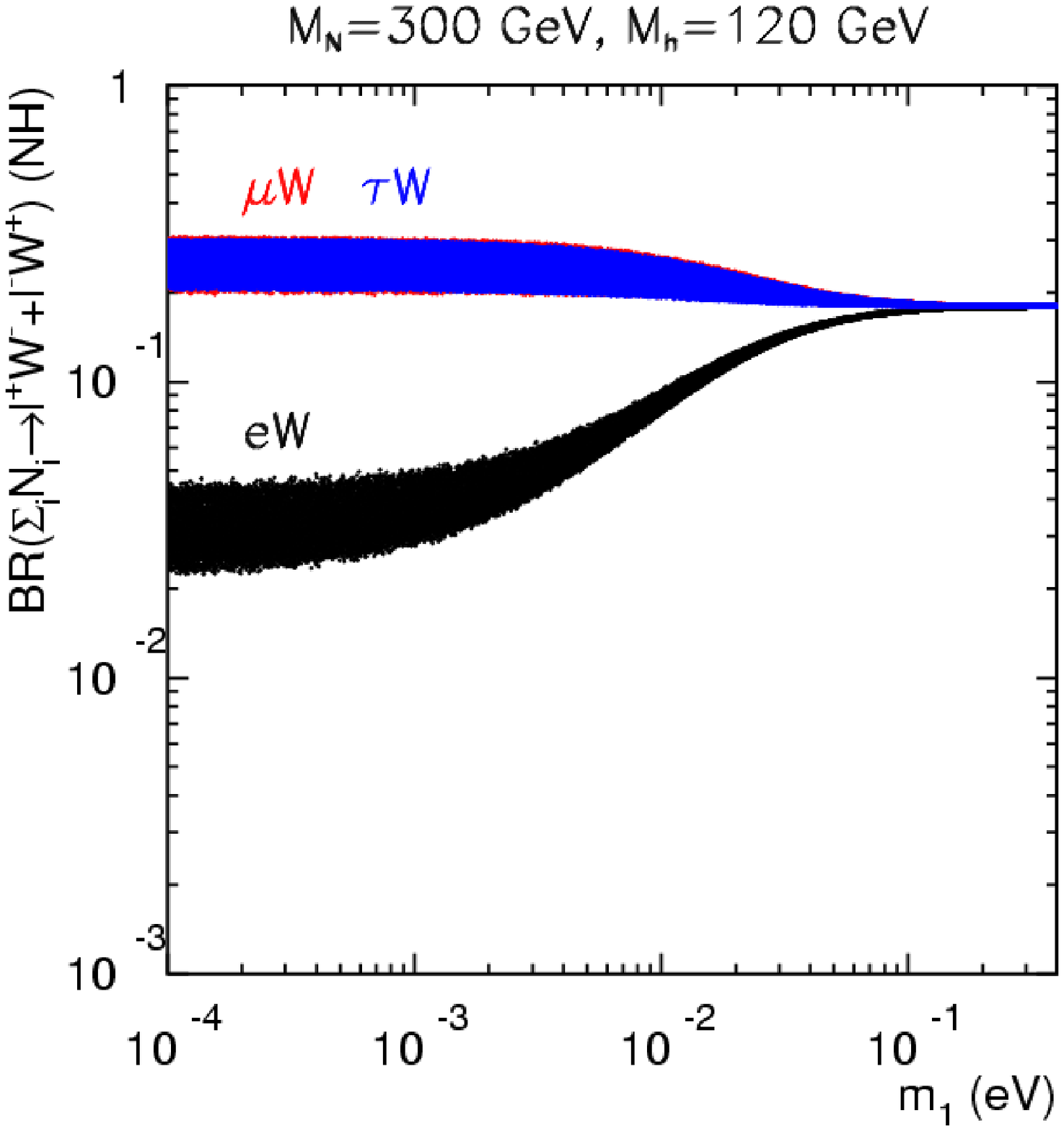}
\includegraphics[scale=1,width=8cm]{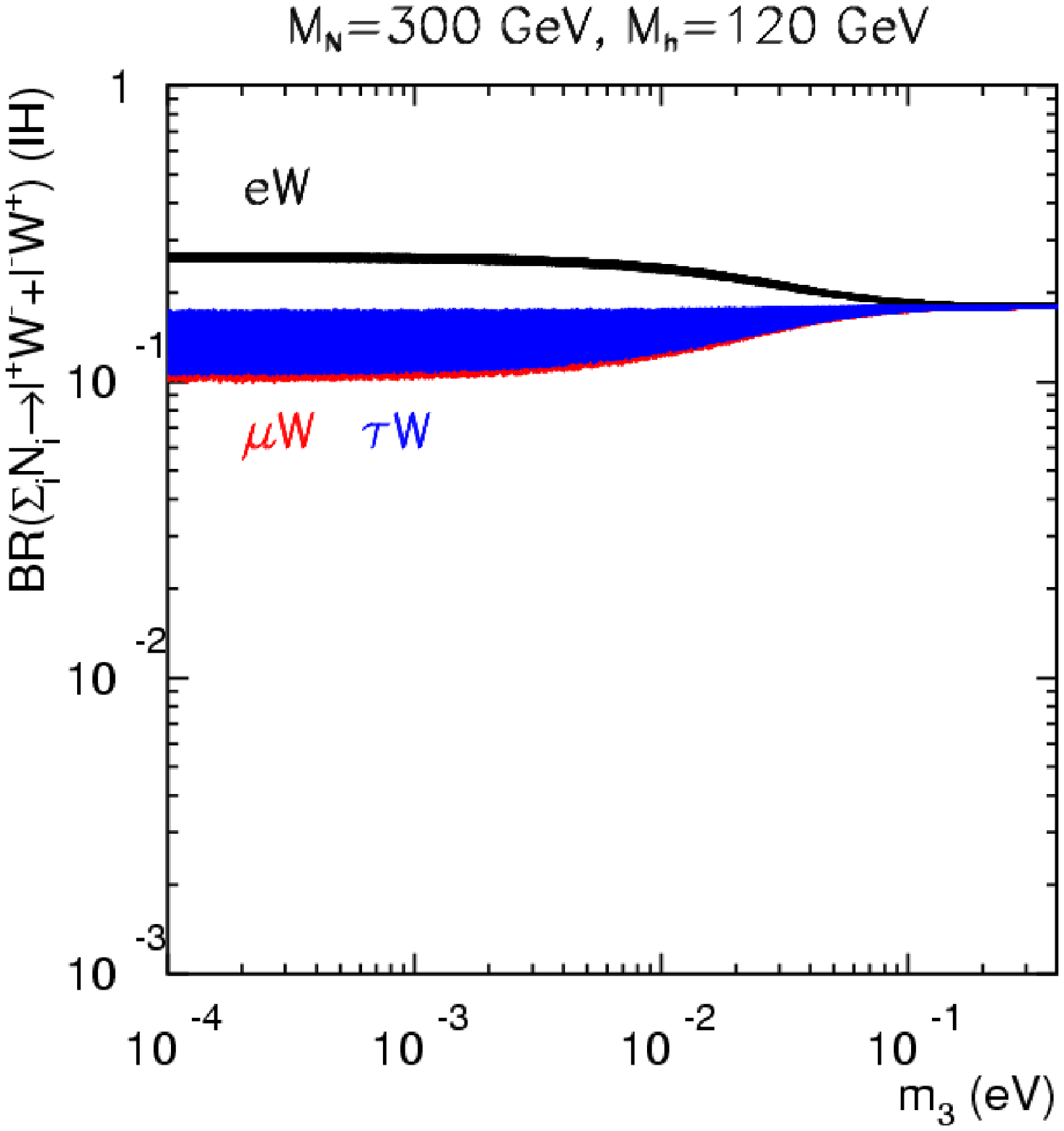}
\end{tabular}
\end{center}
\caption{Branching fractions of degenerate neutrinos $\sum_iN_i \to
\ell^+ W^-+\ell^-W^+ \ (\ell=e,\mu,\tau)$ for NH and IH versus
lightest neutrino mass with $M_N=300~{\rm GeV}$ and $M_{h}=120~{\rm
GeV}$, assuming vanishing Majorana phases.} \label{nbr}
\end{figure}

In Fig.~\ref{nbr} we show the impact of the neutrino masses and
mixing angles on the branching fractions of the sum of the
degenerate neutrinos $N_i \ (i=1,2,3)$ decaying into $e,\mu,\tau$
lepton plus $W$ boson, respectively, with the left panels for the
Normal Hierarchy (NH) and the right panels of the Inverted Hierarchy (IH),
assuming vanishing Majorana phases.
Qualitatively, it follows the relations in Eq.~(\ref{hiera})
\begin{eqnarray}
&& BR(\mu^\pm W^\mp),BR(\tau^\pm W^\mp)\gg BR(e^\pm W^\mp)
\ \ \ {\rm for \ \ NH},\nonumber \\
&& BR(e^\pm W^\mp)>BR(\mu^\pm W^\mp),BR(\tau^\pm W^\mp)
\ \ \ {\rm for \ \ IH}.
\end{eqnarray}
The branching fraction can differ by one order of magnitude in NH case;
and about a factor of few in the IH spectrum. As one
expects that all these channels are quite similar when the neutrino
spectrum is quasi-degenerate, $m_1\approx m_2\approx m_3\geq 0.05$ eV.
Therefore, in this simple case one can hope that if the heavy
neutrino decays are observed in future experiments one should be
able to distinguish the neutrino spectrum.
\subsubsection{\bf Decays in Case II: Non-degenerate Heavy Neutrinos}

\begin{figure}[tb]
\begin{center}
\begin{tabular}{cc}
\includegraphics[scale=1,width=11cm]{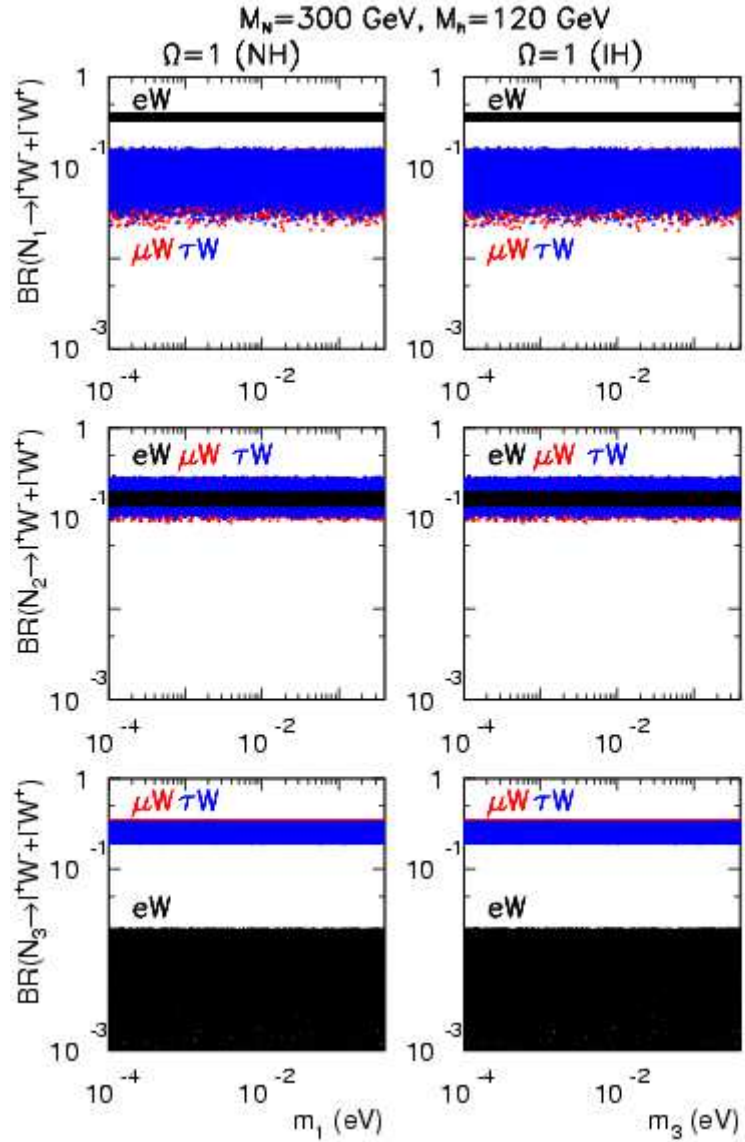}
\end{tabular}
\end{center}
\caption{Branching fractions of process $N_i\to \ell^+ W^-+\ell^-
W^+ (\ell=e,\mu,\tau, i=1,2,3)$ versus the lightest neutrino mass
for NH and IH in Case IIa ($\Omega=I$), when $M_i=300~{\rm GeV}$ and
$M_h=120~{\rm GeV}$. } \label{bri}
\end{figure}

For non-degenerate neutrino spectra we once again study the simple choice:
{\bf Case IIa $\Omega=I$}.
We show the branching fractions of processes $N_i\to \ell^+ W^-+\ell^-W^+ \
(\ell=e,\mu,\tau, \ i=1,2,3)$ corresponding to the
lightest neutrino mass for NH and IH for  $M_i=300~{\rm GeV}$
in Fig.~\ref{bri}. As noted earlier, in this simplest
case all $|V_{\ell i}|^2 \ (\ell=e,\mu,\tau)$ are proportional to
$m_i$. Therefore the branching ratio of $N_i\to \ell^\pm W^\mp$ for each
lepton flavor is independent of neutrino mass and thus universal for both
NH and IH. Although we cannot
distinguish the neutrino mass hierarchy, we still can tell the
difference of the three heavy Majorana neutrinos according to
different SM lepton flavors in final states of their dominant decay
channels. One has
\begin{eqnarray}
&& BR(e^\pm W^\mp) > BR(\mu^\pm W^\mp),\ BR(\tau^\pm W^\mp)
\ \ \ {\rm for}  \ \ N_1,\nonumber \\
&& BR(e^\pm W^\mp)\approx BR(\mu^\pm W^\mp)\approx BR(\tau^\pm W^\mp)
\ \ \ {\rm for}  \ \ N_2,\nonumber \\
&& BR(\mu^\pm W^\mp),\ BR(\tau^\pm W^\mp)\gg BR(e^\pm W^\mp)
\ \ \ {\rm for}  \ \ N_3.\nonumber
\end{eqnarray}
This follows closely to the mixing strengths of the light neutrinos in the previous section.

As discussed previously, {\bf Case IIb} $\Omega = I_{\rm off}$ is
identical to the above if we identify $N_1 \leftrightarrow N_3$.
More involved case for $\Omega$ may be some form of superposition of
the three decay patterns, that is to be tested experimentally by the
flavor combinations.

\begin{figure}[tb]
\begin{center}
\begin{tabular}{cc}
\includegraphics[scale=1,width=8cm]{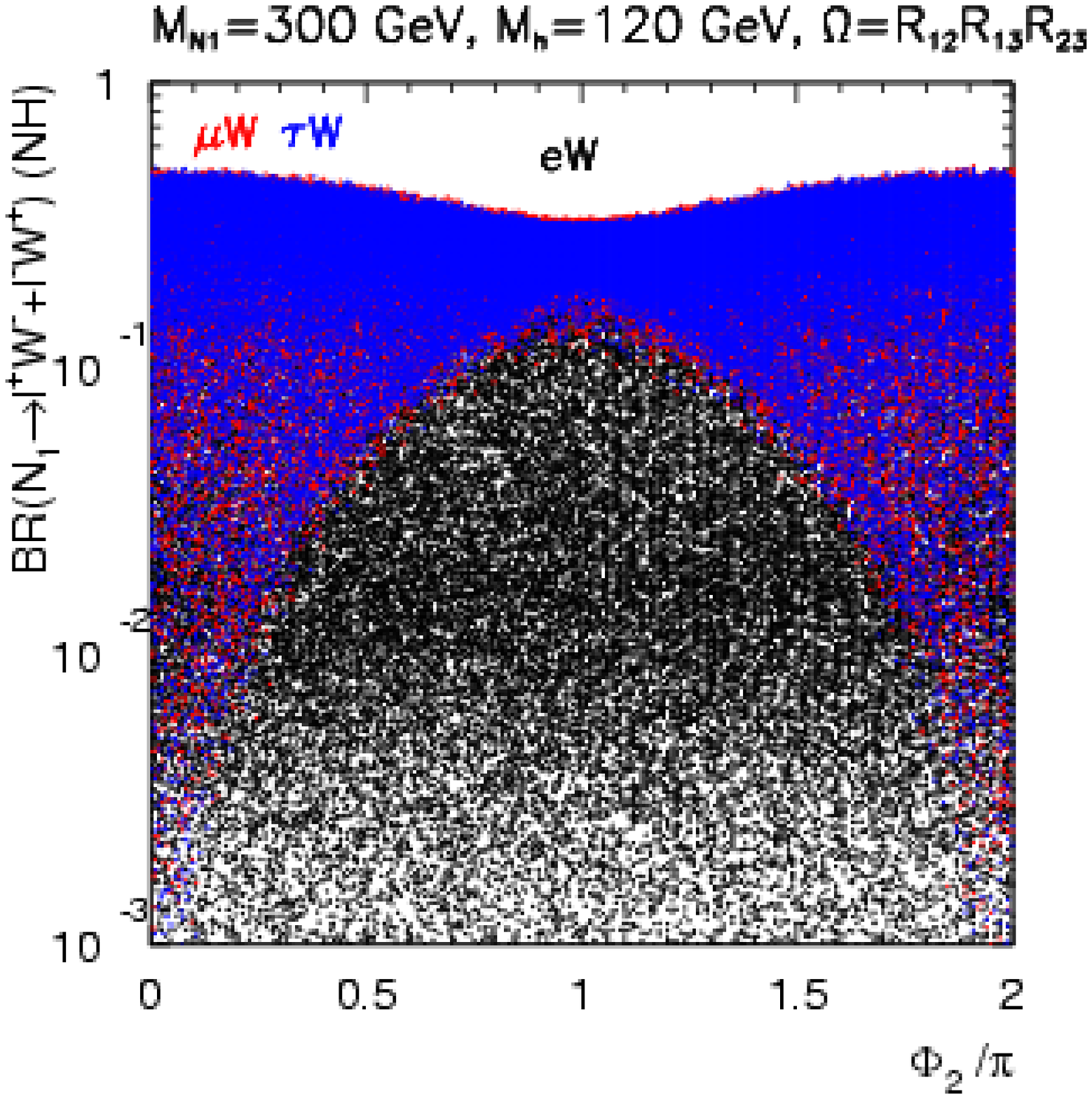}
\includegraphics[scale=1,width=8cm]{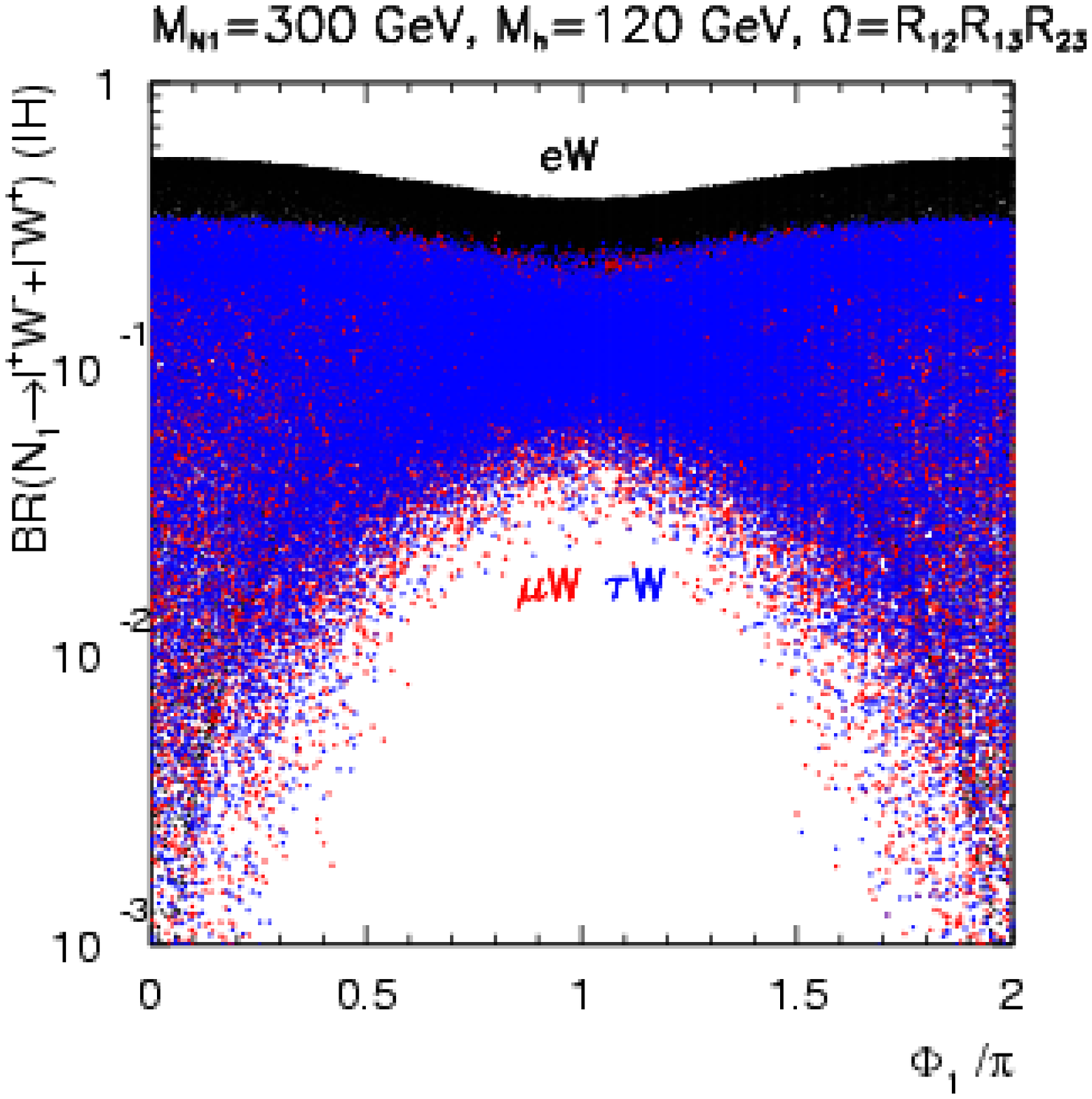}
\end{tabular}
\end{center}
\caption{Branching fractions of $N_1\to \ell^+ W^-+\ell^- W^+$
versus Majorana phase $\Phi_2$ for NH and $\Phi_1$ for IH in the
general non-degenerate case when $M_1=300~{\rm GeV}, M_h=120~{\rm
GeV}$, with random selection of the $\Omega$ matrix elements. }
\label{phi1}
\end{figure}

\subsection{\bf Impact of Majorana Phases in Heavy Majorana Neutrino Decays}

In our previous discussion we have shown that the mixings $|V_{\ell
N}|^2$ are independent of Majorana phases in both Case IIa and IIb
(as well for an $\Omega$ with unity as entries).
In general, the $N_i$ decay rates
depend on only one Majorana phase $\Phi_2\ (\Phi_1)$
when $m_{1(3)}\approx 0$ and $s_{13}=0$ in the NH (IH) case
as shown explicitly in the appendices.
In Fig.~\ref{phi1}, we show the dependence of $N_1$ decay branching
fractions for general non-degenerate case on Majorana phases
$\Phi_2$ and $\Phi_1$ in NH and IH, with random selection of the
$\Omega$ matrix elements.
The dependence
of $N_2$ and $N_3$ decays on Majorana phases are almost the same as
that of $N_1$. The branching fractions of $\mu^\pm W^\mp, \tau^\pm
W^\mp$ ($e^\pm W^\mp$) are typically dominant over all the range of $\Phi_2$
($\Phi_1$) in NH (IH). The dependence on the phases for the leading channels
are rather weak and it is thus hard to extract the phase information from
heavy Majorana neutrino decay. Some typical situations may be similar to the
cases discussed in Ref.~\cite{ourtype2}, and we will not pursue further for the
phase effects.

\subsection{\bf Total Decay Width of Heavy Majorana Neutrino}
\begin{figure}[tb]
\begin{center}
\begin{tabular}{cc}
\includegraphics[scale=1,width=8cm,angle=90]{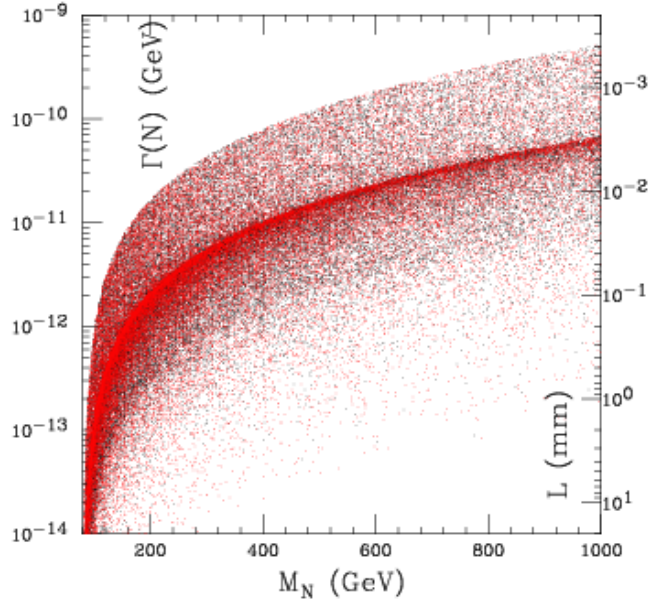}
\end{tabular}
\end{center}
\caption{The total width and decay length of $N$ in the general
non-degenerate case, when the lightest neutrino mass $10^{-4}~{\rm
eV}\leq m_{1(3)}\leq 0.4~{\rm eV}$, $M_{h}=120~{\rm GeV}$ and
$\Omega=R_{12}R_{13}R_{23}$ with random selection of the matrix
elements and .} \label{totw}
\end{figure}

To complete this section about the heavy Majorana neutrino properties,
we study their total decay widths, which are proportional to $M_\nu M_N^2/M_W^2$.
In Fig.~\ref{totw}, we plot the total width (left axis) and decay length (right axis)
for $N$ versus $M_N$ under the general non-degenerate case
with random selection of the $\Omega$ matrix elements  (similar for NH and IH).
There is a large spread for the possible ranges of the
decay lengths, governed by the mixing parameters.
Although not generally considered as long-lived for large mass,
the $N$ decay lengths may be typically in the range of $\mu$m$-$cm,
and their decays could lead to a visible displaced vertex in the detector at the LHC.

\begin{figure}[tb]
\begin{center}
\begin{tabular}{cc}
\includegraphics[scale=15,width=12cm]{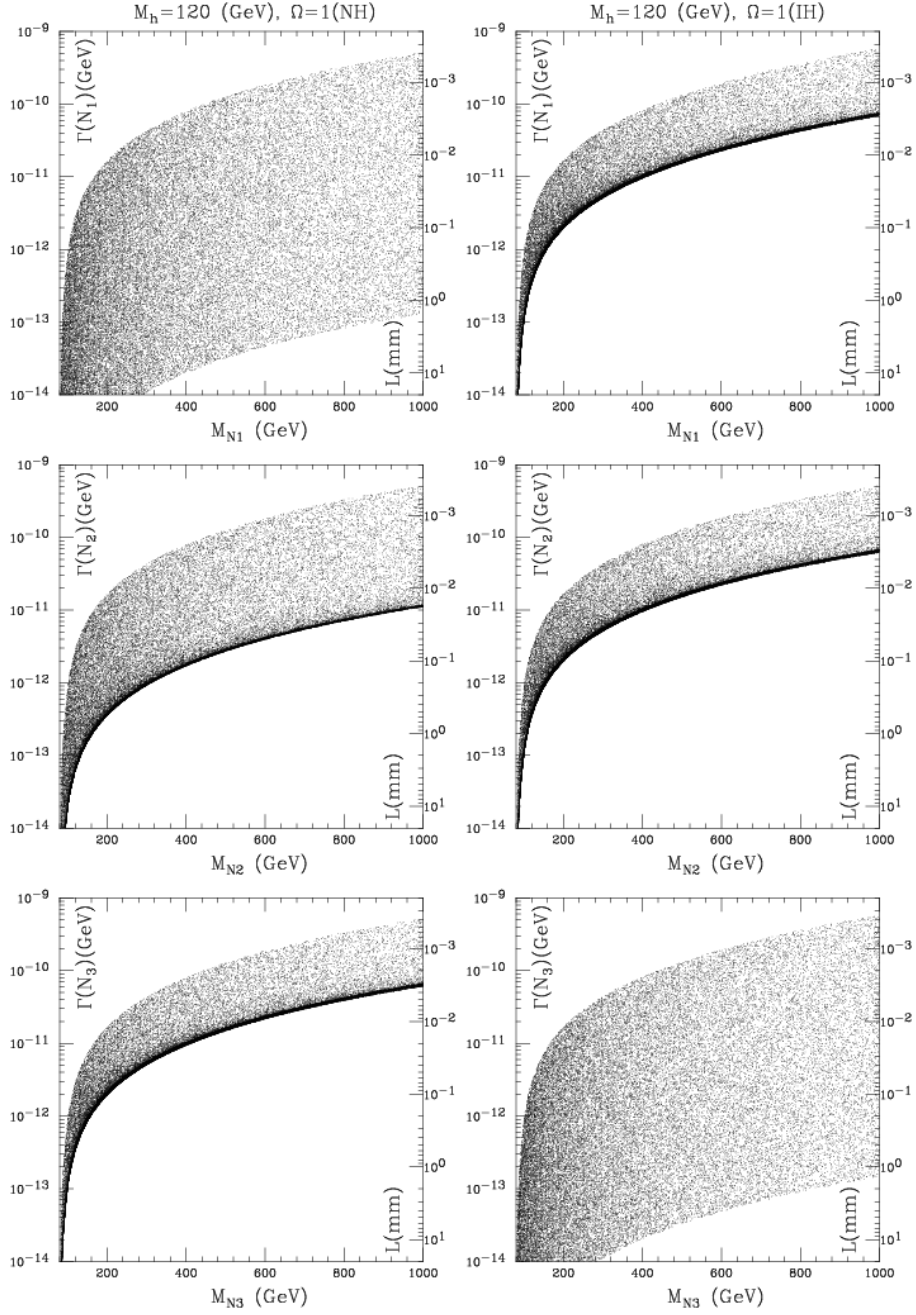}
\end{tabular}
\end{center}
\caption{Total width and decay length of $N_i (i=1,2,3)$ for NH and
IH in Case IIa ($\Omega=I$), when the lightest neutrino mass
$10^{-4}~{\rm eV}\leq m_{1(3)}\leq 0.4~{\rm eV}$ and $M_{h}=120~{\rm
GeV}$.} \label{totwd1}
\end{figure}

When considering a specific model-parameter setting,
we plot the total width (left axis) and decay length (right axis) in Fig.~\ref{totwd1},
for $N_i$ versus $M_N$ for $M_h=120~{\rm GeV}$
in NH and IH under Case IIa with $\Omega=I$.
One of the generic features for all $N_i$ and both NH and IH is a typical lower limit
for their lifetime (or  decay length). For instance, the typical decay length
for $M_N\gtrsim 600~{\rm GeV}$ is above $1~{\rm \mu m}$.
For smaller values of $M_N$, the heavy Majorana
neutrinos can be long-lived in the detector scale, making the
signatures detectable at the secondary vertex. In fact, this feature remains
in a majority part of the parameter space.
In particular,
because in this case all $|V_{\ell i}|^2$ are proportional to $m_i$,
the lifetimes of $N_1$ in NH and $N_3$ in IH could be infinite when
neglecting the lightest neutrino mass in whole Majorana neutrino
mass range.
It is interesting to note that  there is a clear difference between the NH and IH scenarios:
the lifetime of $N_1$ in IH and $N_3$ in NH has a narrowly predicted range within one
order of magnitude, about 10 $\mu$m for $M_N=400$ GeV.
If this is indeed observed, it could serve as an indication to distinguish the models.
The lifetimes of $N_2$ in NH and IH are almost the same.

For Case IIb with an off-diagonal $\Omega$ matrix,
the lifetime features of $N_1$ and $N_3$ are also
interchanged with each other and those of $N_2$ are still the same as
Case IIa.

\section{Heavy Majorana Neutrinos and the test of Type I seesaw at the LHC}
In order to study the prediction for the lepton flavor correlations
with heavy Majorana neutrino and its LNV decay processes, the ideal
production channels are the Drell-Yan processes
via SM gauge bosons, $pp\to W\to N\ell,\ pp\to Z\to NN$.
However, the gauge couplings to  $N$ are highly suppressed to the order
$\mathcal{O}(m_\nu/M_N)$~\cite{taotypei}. The situation is very
different in the case of the minimal $B-L$ extension of the SM (see
appendix C) where one can produce the heavy neutrinos through
the $Z'$ in the theory.
\subsection{Gauge Boson Properties: $Z'$}
In the limit where there is no mixing between the two Abelian sector
of the minimal $B-L$ extension of the SM (see appendix C,
$\epsilon=0$ in Eq.~(\ref{Kinetic})), the mass of the new gauge
boson $Z'$ is given by
\begin{equation}
M_{Z'}= 2 g_{BL}^{} v_S.
\end{equation}
To satisfy the experimental lower bound, $M_{Z'}/g_{BL} > (5-10) $ TeV, it is sufficient
to assume that $v_S > 2.5 - 5 $ TeV.  The relevant interactions to matter are given by
\begin{equation}
g_{BL} \ Z_{\mu}' \ \left( Q_{BL}^q \left[ \bar{u} \gamma^\mu u \
+ \ \bar{d} \gamma^{\mu} d \right] + Q_{BL}^\ell \left[
 \bar{e} \gamma^\mu e + \bar{\nu}_L \gamma^\mu \nu_L +
 \bar{\nu}_R \gamma^\mu \nu_R \right] \right) ,
\end{equation}
where the $B-L$ charges are assigned to be $Q_{BL}^q = 1/3$,  and $Q_{BL}^\ell = -1$.

There has been a lot of work on the heavy neutral gauge bosons. For
a recent review, see Ref.~\cite{Langacker:2008yv}, and recent
studies of $Z'$ at the Tevatron and LHC~\cite{Petriello:2008zr}.
For a recent consideration of the phenomenological aspects of the
$B-L$ model, see~\cite{Huitu}.
\begin{figure}[tb]
\begin{center}
\begin{tabular}{cc}
\includegraphics[scale=1,width=8cm]{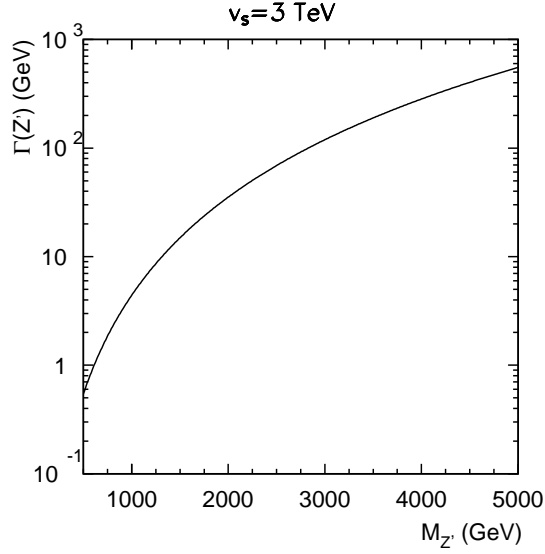}
\end{tabular}
\end{center}
\caption{Total decay width of $Z'$, when $v_S=3~{\rm TeV}$.}
\label{zpw}
\end{figure}
The expressions for the possible decays of the $Z'$ are given by
\begin{eqnarray}
\Gamma(Z'\to f\bar{f})&=&g_{BL}^2 {M_{Z'}\over 12\pi} C_f\
(Q_{BL}^f)^2
\left(1+2{m_f^2\over M_{Z'}^2}\right)\ \beta_f , \\
\Gamma(Z'\to \sum_m \nu_m \nu_m)&=& 3 g_{BL}^2 {M_{Z'}\over
24\pi}C_\nu (Q_{BL}^\ell)^2,
\\
\Gamma(Z'\to N_m N_m)&=&g_{BL}^2 {M_{Z'}\over 24\pi}C_N
(Q_{BL}^\ell)^2 \ \beta_N^{3} .
\end{eqnarray}
where $f=\ell,q$, the couplings $C_{\ell,\nu,N}=1,C_q=3$,
and $\beta_i = \sqrt{1- 4m_i^2/ M_{Z'}^2}$ is the speed of particle $i$. Note that the decay
width to Majorana particles is of a threshold behavior $\beta^3$, and is half of that for
a Dirac particle.
Well above the threshold, the $Z'$ decay branching fractions take the simple ratios for the final states
\begin{equation}
\sum_\ell^{e,\mu,\tau} \ell^+ \ell^- : \sum_q^{u...t} q\bar q :
\sum_m^{1,2,3} \nu_m \nu_m : N_1 N_1 = 3: 2: {3\over 2} : {1\over
2}.
\end{equation}
We show in Fig.~\ref{zpw} the results for the case $v_S=3$ TeV. It scales as
\begin{equation}
\Gamma(tot) \approx 0.2\   g_{BL}^2 \ M_{Z'}   < 0.05 \
\left({M_{Z'}\over v_S}\right)^2 \ M_{Z'}.
\end{equation}
Notice that this $Z'$ has the property that its coupling to quarks
is suppressed with respect to the couplings to leptons. As it well
known the $Z'$ in Left-Right symmetric theories has different
properties from the $B-L$ case studied here. Then, from the standard
analysis where one uses the leptonic channels and the channels into
heavy quarks one can distinguish the $B-L$ case from the rest easily.

It is important to emphasize that in the case of the $B-L$ SM, one
gets an upper bound on the mass of the heavy neutrinos $M_N \leq
M_{Z'}/(2\sqrt{2}g_{BL})$ (see appendix C for details).

\subsection{Heavy Majorana Neutrino Production through $Z'$ mediation at the LHC}
\begin{figure}[tb]
\begin{center}
\begin{tabular}{cc}
\includegraphics[scale=1,width=8cm]{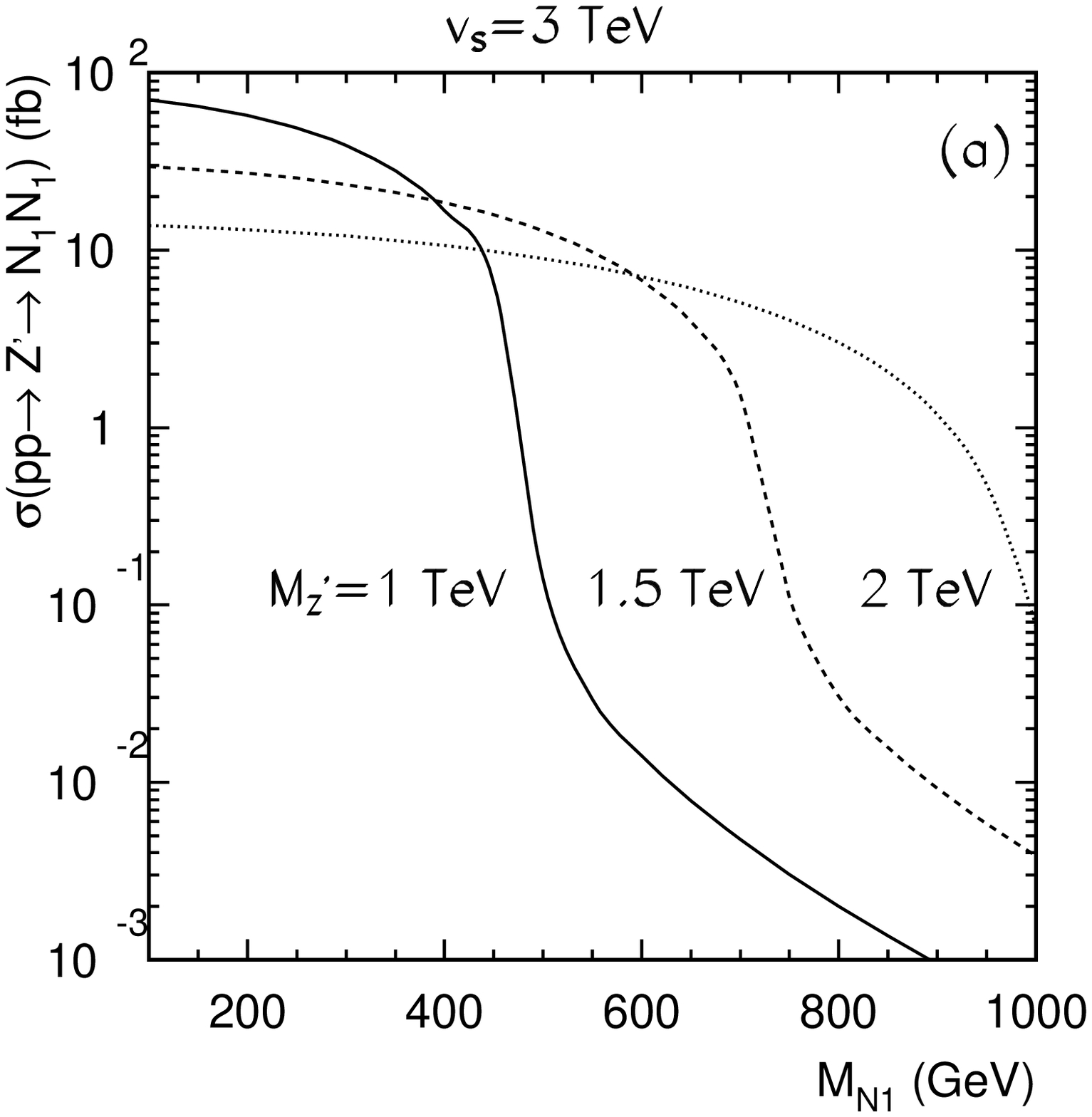}
\includegraphics[scale=1,width=8cm]{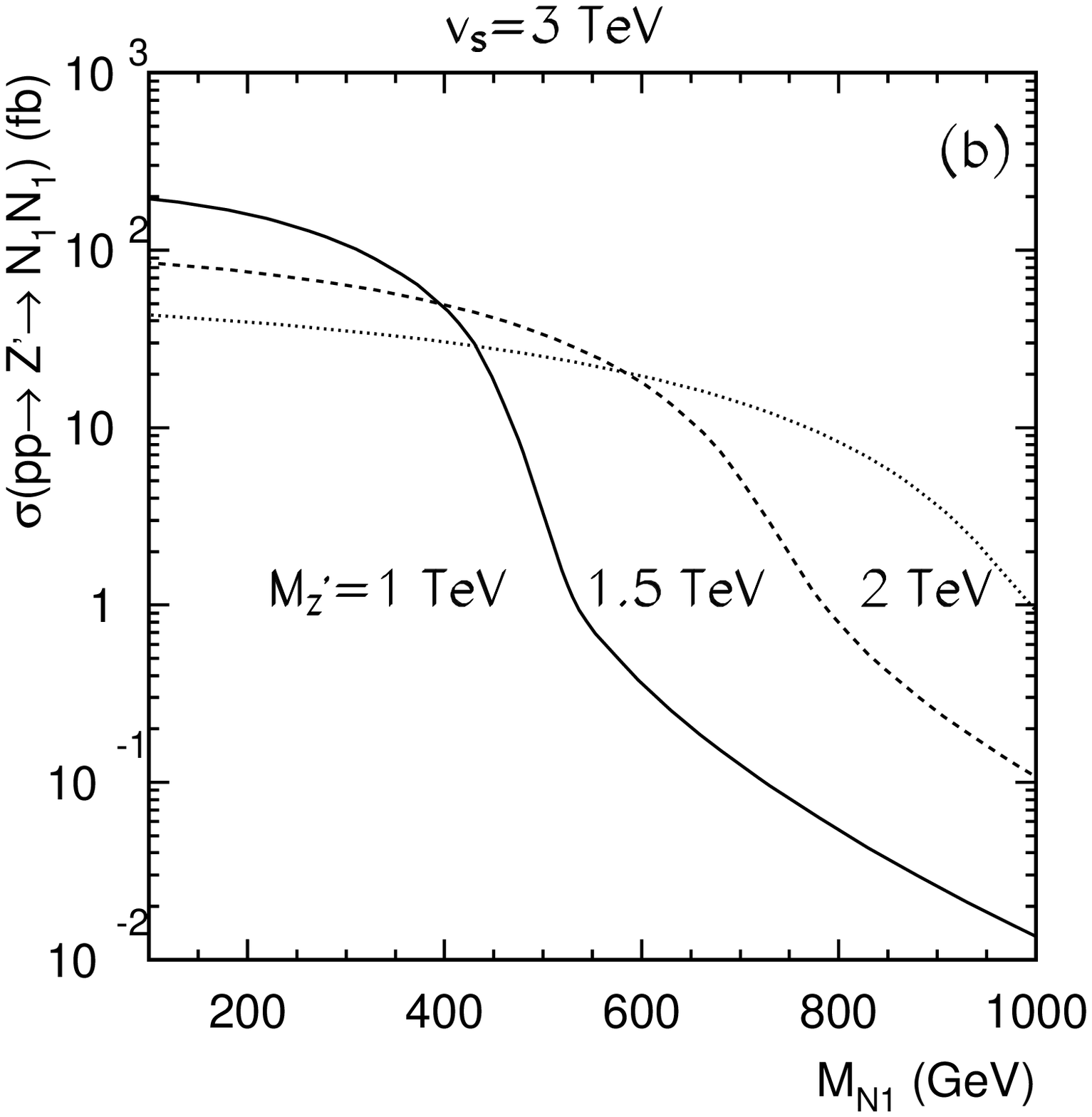}
\end{tabular}
\end{center}
\caption{Heavy Majorana neutrino pair production total cross section
at the LHC versus its mass. The solid, dashed and dotted curves are
for $M_{Z'}=1,1.5,2~{\rm TeV}$ respectively, when $v_S=3~{\rm TeV}$,
(a) for $U(1)_{B-L}$ coupling and (b) for $U(1)_X$ coupling, as
given in Tables~\ref{int} and \ref{int2}. } \label{ncs}
\end{figure}

We are interested in the production of two heavy neutrinos.
Since in this model one has a dynamical mechanism for $B-L$ breaking,
there is a production mechanism through the $Z'$. Then, we are interested in the mechanism
\begin{equation}
p p \to Z' \to N_1 N_1.
\end{equation}
The parton level cross section for this process is
\begin{eqnarray}
&&{d\sigma(q\bar{q}\to Z'\to N_1N_1)\over dt}={1\over 32\pi
s^2N_c}{2g_{BL}^4\over 9}{1\over
(s-M_{Z'}^2)^2+M_{Z'}^2\Gamma^2_{Z'}}\left[(t-M_N^2)^2+(u-M_N^2)^2-2sM_N^2\right]\nonumber\\
\end{eqnarray}
where $t=(p_q-p_N)^2$. The total cross section versus heavy Majorana
neutrino mass at the LHC is plotted in Fig.~\ref{ncs}, assuming
$v_S=3~{\rm TeV}$ with (a) for $U(1)_{B-L}$ coupling and (b) for
$U(1)_X$ coupling, as given in Tables~\ref{int} and \ref{int2}.
We see that the production cross sections are quite sizable, typically of the order of $10-100$ fb.
The cross section drops sharply after reaching the kinematical threshold $2M_N>M_{Z'}$.

The Majorana signals for $\Delta L=2$ decay of $N_1$ are
\begin{eqnarray}
N_1 N_1\to \ell^\pm \ell^\pm\ W^\mp W^\mp , \ \ \ \ell=e,\mu,\tau
\end{eqnarray}
To confirm the important feature of lepton number violation, we demand the $W$'s decay hadronically.
The overall branching fraction to be included becomes
\begin{eqnarray}
{\rm BR}(N_1 N_1\to \ell^\pm \ell^\pm\ 4~jets) \approx 2 \cdot ({1\over 4})^2\cdot ({6\over 9})^2 = {1\over 18}.
\end{eqnarray}
Note that there are also accompanying  clean channels like
$\ell^\pm \ell^\mp + 4~jets $, that are not lepton-number violating and we do not include for the rest of the analysis.

We would like to reiterate that in a significant range of the
parameter space of $M_N$ and mixings, the $N$ decay could lead to
distinctive signatures with a decay length longer than 10 $\mu$m,
resulting in secondary displaced vertices. This may yield
essentially background-free signal for $N$'s. Nevertheless, we now
explore the signal observability according to the different lepton
flavors without relying on the displaced vertex considerations.

For our numerical analyses, we adopt the CTEQ6L1 parton distribution
function~\cite{CTEQ}. We evaluate the SM backgrounds by using the
automatic package Madgraph~\cite{Madgraph}. We work in the
parton-level, but simulate the detector effects by the kinematical
acceptance and employ the Gaussian smearing for the electromagnetic
and hadronic energies~\cite{cms}.

\subsection{$N_1N_1\to \ell^\pm \ell^\pm + \4j \ (\ell=e,\mu)$}
\begin{figure}[tb]
\begin{center}
\begin{tabular}{cc}
\includegraphics[scale=1,width=8cm]{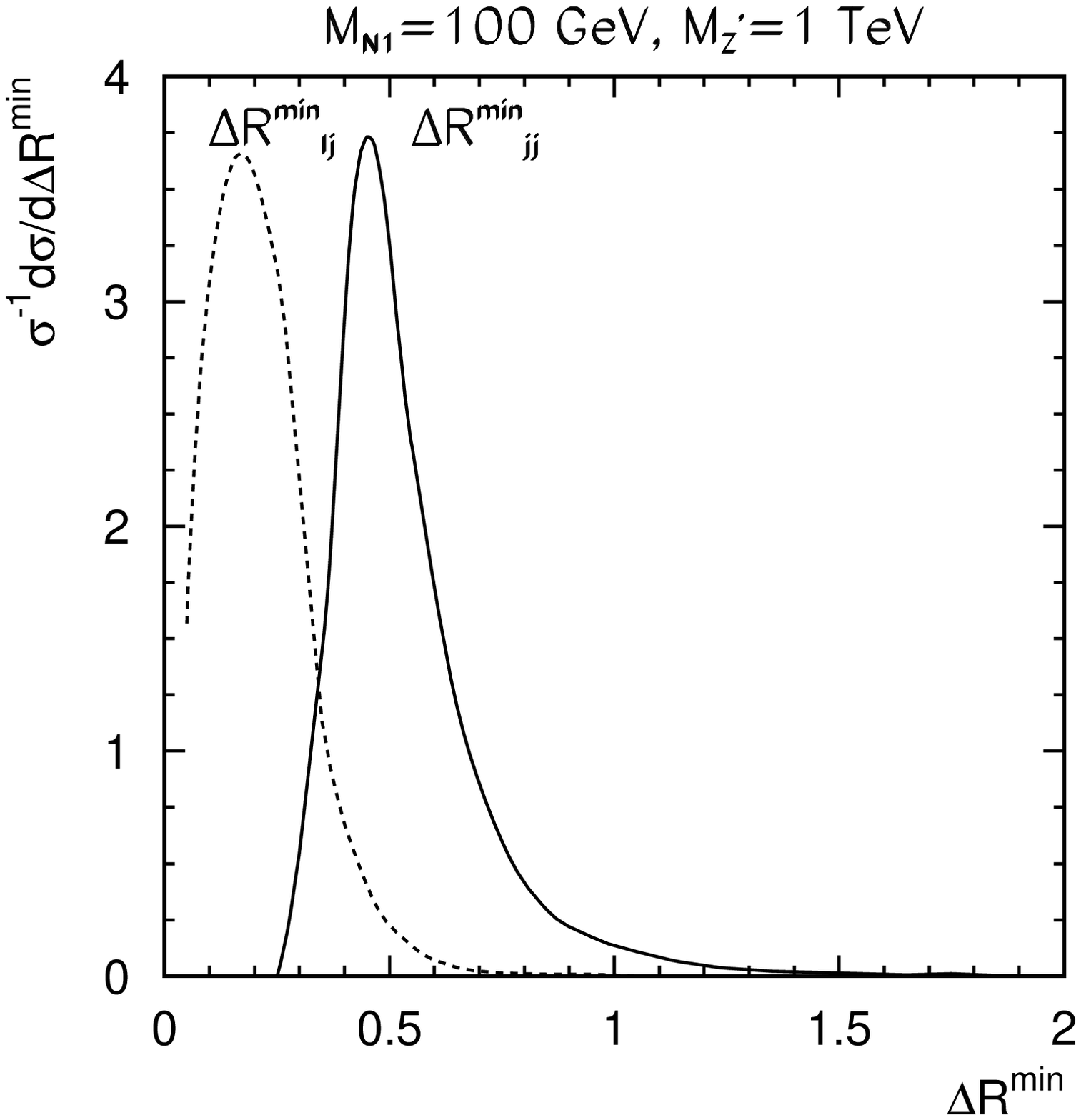}
\includegraphics[scale=1,width=8cm]{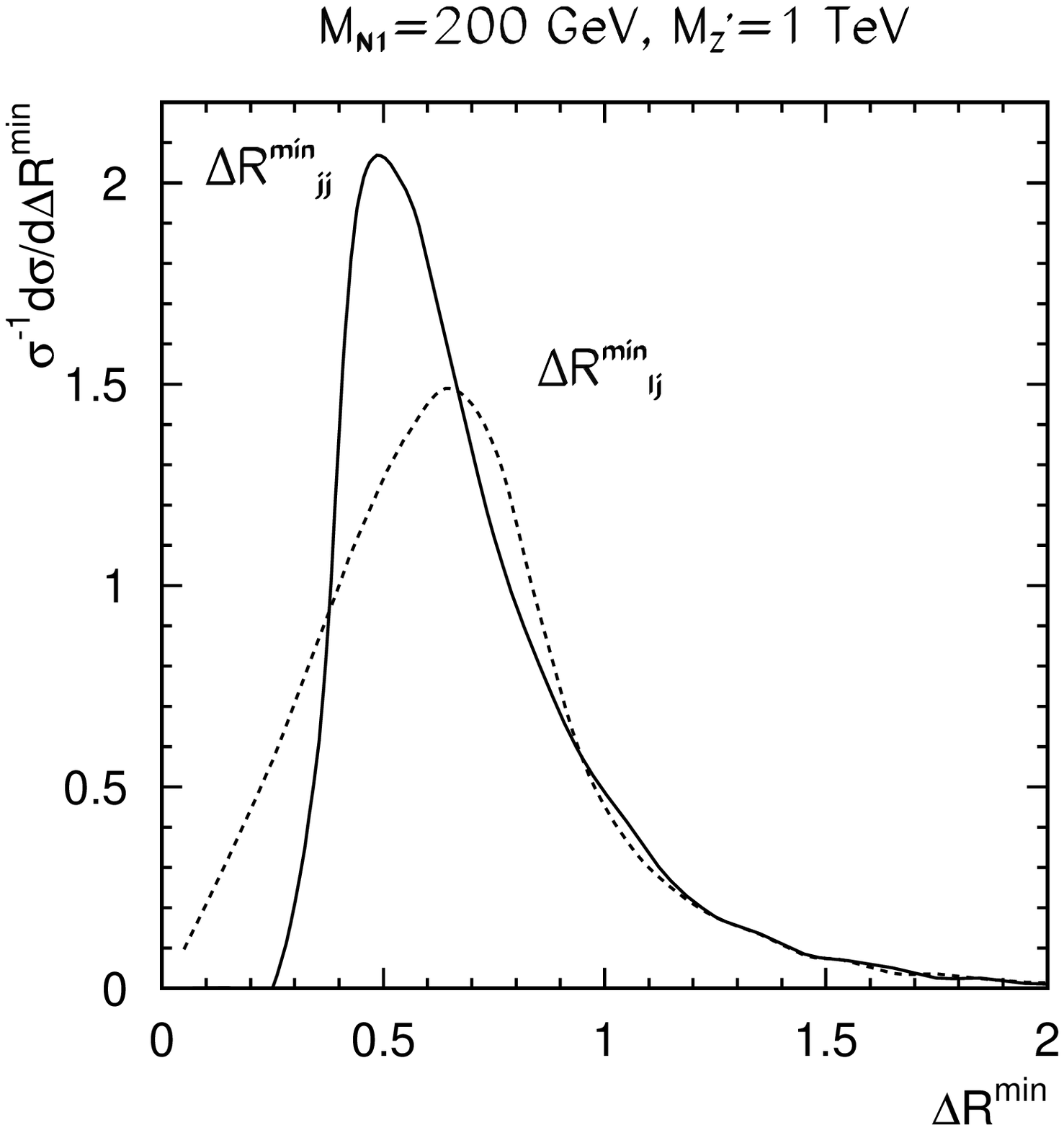}
\end{tabular}
\end{center}
\caption{$\Delta R^{min}_{jj}$ and $\Delta R^{min}_{\ell j}$ for
$M_{N_1}=100~{\rm GeV}$ (left) and $M_{N_1}=200~{\rm GeV}$ (right),
with $M_{Z'}=1~{\rm TeV}$.} \label{zpnr1}
\end{figure}

We start from the cleanest channels with $e,\mu$ in the final state from $N_1$ decay.
We employ the following basic acceptance cuts for the event selection~\cite{cms}
\begin{eqnarray}
&&p_T(\ell)\geq15~{\rm GeV}, \ |\eta(\ell)|<2.5,\\
&&p_T(j)\geq25~{\rm GeV}, \ |\eta(j)|<3.0, \\
&&\Delta R_{jj}\geq 0.3,\ \Delta R_{j\ell},\ \Delta R_{\ell\ell}\geq 0.4
\end{eqnarray}
The rather loose cuts on the separations $\Delta R$ are designed to keep the signal events for a heavier $Z'$
and a lighter $N$ which is fast moving and thus yields collimated decay products of a lepton and two jets.
We plot the minimal isolation $\Delta R^{min}_{jj}$ of two jets and $\Delta R^{min}_{\ell j}$ of one jet and one charged lepton for $M_{Z'}=1~{\rm TeV}$ and $M_N=100,\ 200$ GeV, respectively,  in Fig.~\ref{zpnr1}.
One can see that for $M_{N}\gtrsim 200~{\rm GeV}$ with $M_{Z'}=1~{\rm TeV}$
the signal consists of well-isolated one pair of same-sign leptons of arbitrary
$e,\mu$ flavor combinations plus four light jets.

To simulate the detector effects on the energy-momentum measurements,
we smear the electromagnetic energy, the electromagnetic energy and jet energy
by a Gaussian distribution whose width is parameterized as~\cite{cms}
\begin{eqnarray}
{ \Delta E\over E} &=& {a_{cal} \over \sqrt{E/{\rm GeV}} } \oplus b_{cal}, \quad
a_{cal}=10\%,\  b_{cal}=0.7\% ,
\label{ecal}\\
{ \Delta E\over E} &=& {a_{had} \over \sqrt{E/{\rm GeV}} } \oplus b_{had}, \quad
a_{had}=50\%,\  b_{had}=3\%.
\end{eqnarray}

\begin{figure}[tb]
\begin{center}
\begin{tabular}{cc}
\includegraphics[scale=1,width=9cm]{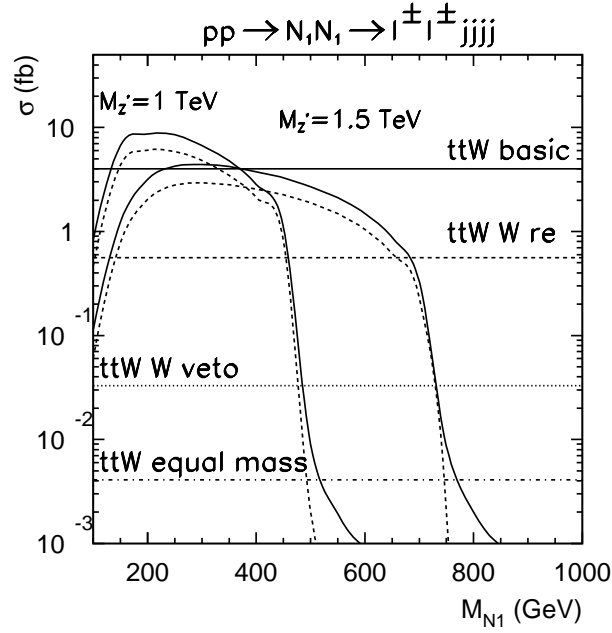}
\end{tabular}
\end{center}
\caption{Production cross section of $N_1N_1$ with basic cuts and
hard final states cuts. Branching fractions for $N_1$ decay to charged leptons
are not included; while $W$ decays to 2 jets are included.
For comparison, the background process is
also included with the sequential cuts as indicated.}
\label{zpnpair}
\end{figure}

In principle, there is no genuine SM background to the lepton-number violating processes.
The leading SM  background to our signal is from  decays of two like-sign $W$'s to leptons.
For instance, the leading reducible background to our signal is
\begin{eqnarray}
pp\to t\bar{t}W^\pm\to W^\pm W^\pm jjb\bar{b} .
\end{eqnarray}
The QCD processes $jjjjW^\pm W^\pm, jjW^\pm W^\pm W^\mp$ are much
smaller. This is estimated based on the fact that QCD $jjW^\pm
W^\pm\rightarrow jj\ell^\pm\ell^\pm\cancel{E}_T$ is about 15 fb.
With an additional $\alpha^2_s$ and 6 body phase space or one more
$W$ suppression, they are much smaller than $t\bar{t}W^\pm$. Other
EW backgrounds $WWWW,WWWZ$ are also neglectable. Although the
background rates are large to begin with, the kinematics is quite
different between the signal and the backgrounds. We outline the
characteristics and propose some judicious cuts as follows.
\begin{itemize}
\item The SM backgrounds always come with $W$ pair decays with missing neutrinos.
To suppress backgrounds, we veto the events with large missing
energy $\cancel{E}_T<20~{\rm GeV}$.

\item We choose the two pairs with nearly equal masses from the six dijet
combinations as the two hadronic $W$'s and take $W$ boson
reconstruction as $|M_{jj}-M_W|<15~{\rm GeV}$. The efficiency is
very high.

\item In order to select the correct lepton and two jets combination and reconstruct $N_1$,
we take advantage of the feature that the two heavy neutrinos have
equal masses $M_{\ell_1 j_1j_2}=M_{\ell_2 j_3 j_4}$. In practice, we
impose $|M_{\ell_1 j_1j_2}-M_{\ell_2 j_3 j_4}|<M_{N_1}/25$. This
helps for the background reduction.

\end{itemize}
The production cross section of $N_1N_1$ signal with the basic cuts
(solid curve) and all of the cuts above (dashed curve) are plotted
in Fig.~\ref{zpnpair}, where
branching fractions for $N_1$ decay to charged leptons
are not included; while $W$ decays to 2 jets are included.
For comparison, the background process of
$t\bar{t}W^\pm$ is also included with the sequential cuts as
indicated. The background is suppressed substantially.

When performing the signal significance analysis, we look for the
resonance in the mass distributions of $\ell jj$ and $2\ell 4j$.
If we look at mass window of
$|M_{\ell_1j_1j_2,\ell_2j_3j_4}-M_{N_1}|<M_{N_1}/20$ and
$|M_{2\ell+4j}-M_{Z'}|<M_{Z'}/30$, the background will be at a
negligible level.

\subsection{$N_1N_1\to \tau^\pm \ell^\pm + \4j$}

The previous section sets the stage for the analyses in the following
sections. Most of the issues for event selection and detector acceptance
will remain the same for the following studies. The next presentations will thus
be sketchy and mainly outlining the new features, in particular the $\tau$-reconstruction
and the mass resonances.

The $\tau$ lepton final state from heavy Majorana neutrino decay
plays an important role in distinguishing different neutrino mass
patterns. Its identification and reconstruction are different from
$e,\mu$ final states because a $\tau$ decays promptly and there will
always be missing neutrinos in $\tau$ decay products.
In practice when selecting events with $\tau$'s, we
require a minimal missing transverse energy
\begin{equation}
\cancel{E}_T > 20\ {\rm GeV}.
\end{equation}
This will effectively separate them from the $\ell\ell jjjj$ type of signal events.

\begin{figure}[tb]
\begin{center}
\begin{tabular}{cc}
\includegraphics[scale=1,width=8cm]{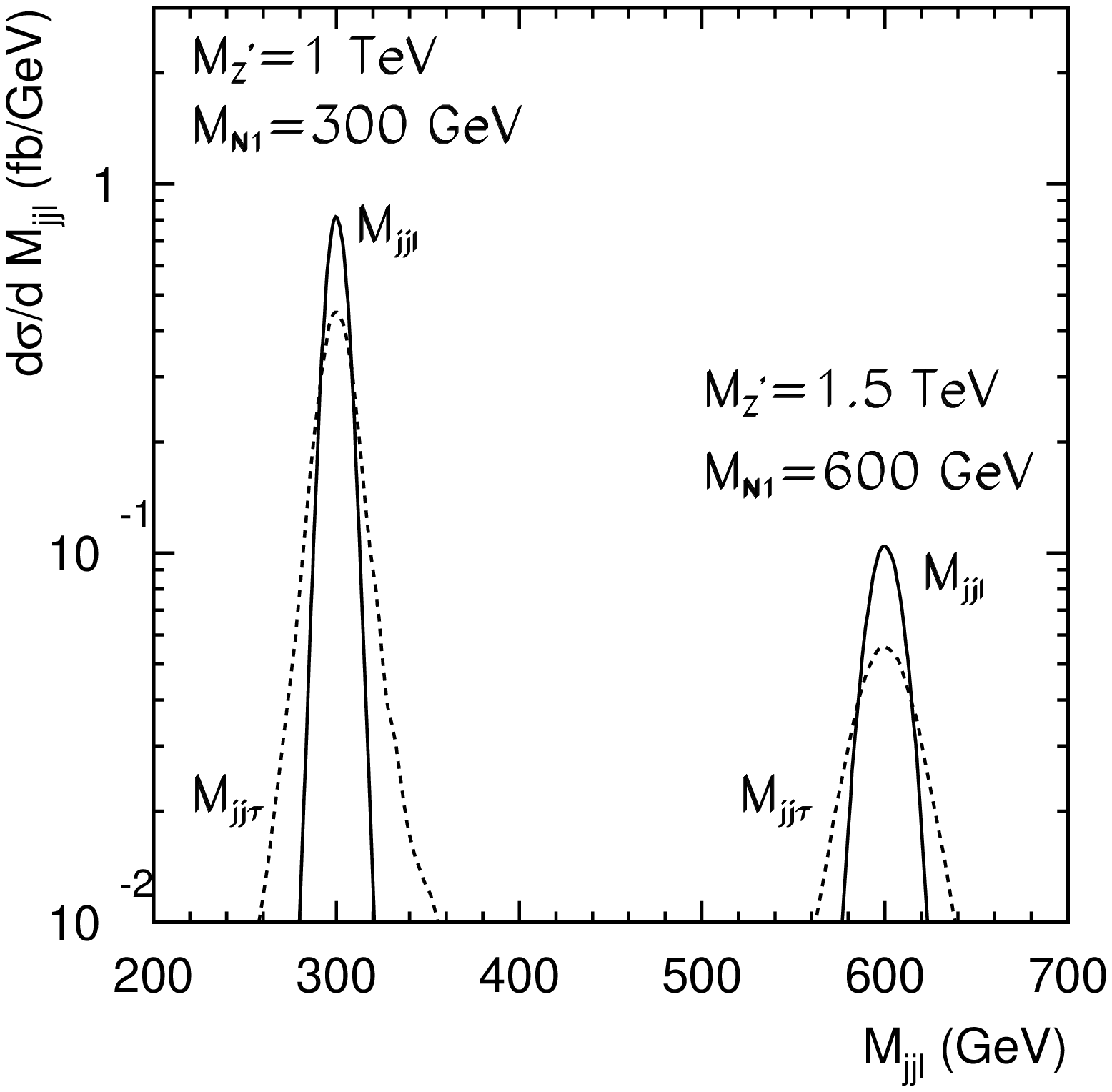}
\includegraphics[scale=1,width=8cm]{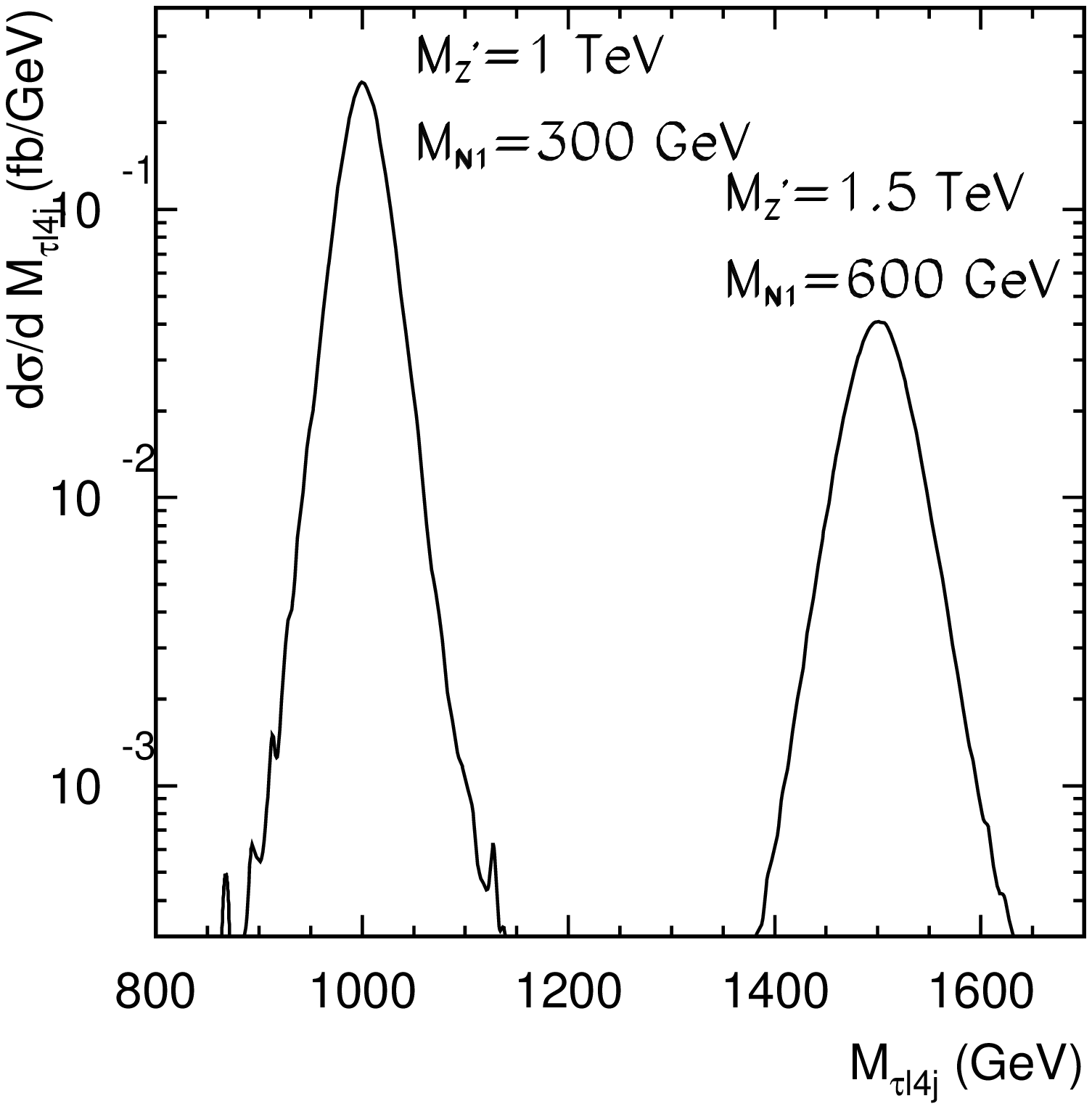}
\end{tabular}
\end{center}
\caption{Reconstructed invariant mass of $M(jj\ell),\ M(jj\tau)$
for $M_{N_1}=300,\  600~{\rm GeV}$, respectively (left), and $M(\tau\ell 4j)$
for $M_{Z'}=1,\  1.5~{\rm TeV}$ (right).}
\label{ren1t}
\end{figure}

We first note that all the $\tau$'s are very energetic from the
decay of a few hundred ${\rm GeV}$ $N_1$. The missing momentum will
be along the direction of the charged track. We thus assume the
momentum of  the missing neutrinos to be reconstructed by
\begin{eqnarray}
\overrightarrow{p}({\rm invisible}) =\kappa\overrightarrow{p}({\rm
track}).
\end{eqnarray}
Identifying $\overrightarrow{p_T}({\rm invisible})$ with the
measured $\cancel{E}_T$, we thus obtain the $\tau$ momentum by
\begin{eqnarray}
\nonumber \overrightarrow{p}_T(\tau)=\overrightarrow{p}_T(\ell)
+\overrightarrow{\cancel{E}}_T,\quad p_L^{}(\tau)=p_L^{}(\ell) + {
\cancel{E}_T \over p_T^{}(\ell) } p_L^{}(\ell).
\end{eqnarray}
The $N_1$ pair kinematics is thus fully reconstructed. The
reconstructed invariant masses of $M(\ell jj)$ and $M(\tau jj)$ are
plotted in Fig.~\ref{ren1t}. We see that $M(\tau jj)$ distribution (dotted curve)
is slightly broader as anticipated. The rather narrow mass peak of
the $\ell jj$ system nevertheless serves as the most distinctive
kinematical feature for the signal identification. Invariant masses
of $M(\tau \ell+4j)$ are also plotted in Fig.~\ref{ren1t}.
Although the existence of missing energy in the signal  makes the background
separation more involved, the resonant mass reconstruction proves to be high
efficient and the backgrounds can still be suppressed to a negligible level.

\subsection{$N_1N_1\to \tau^\pm \tau^\pm + \4j$}

\begin{figure}[tb]
\begin{center}
\begin{tabular}{cc}
\includegraphics[scale=1,width=8cm]{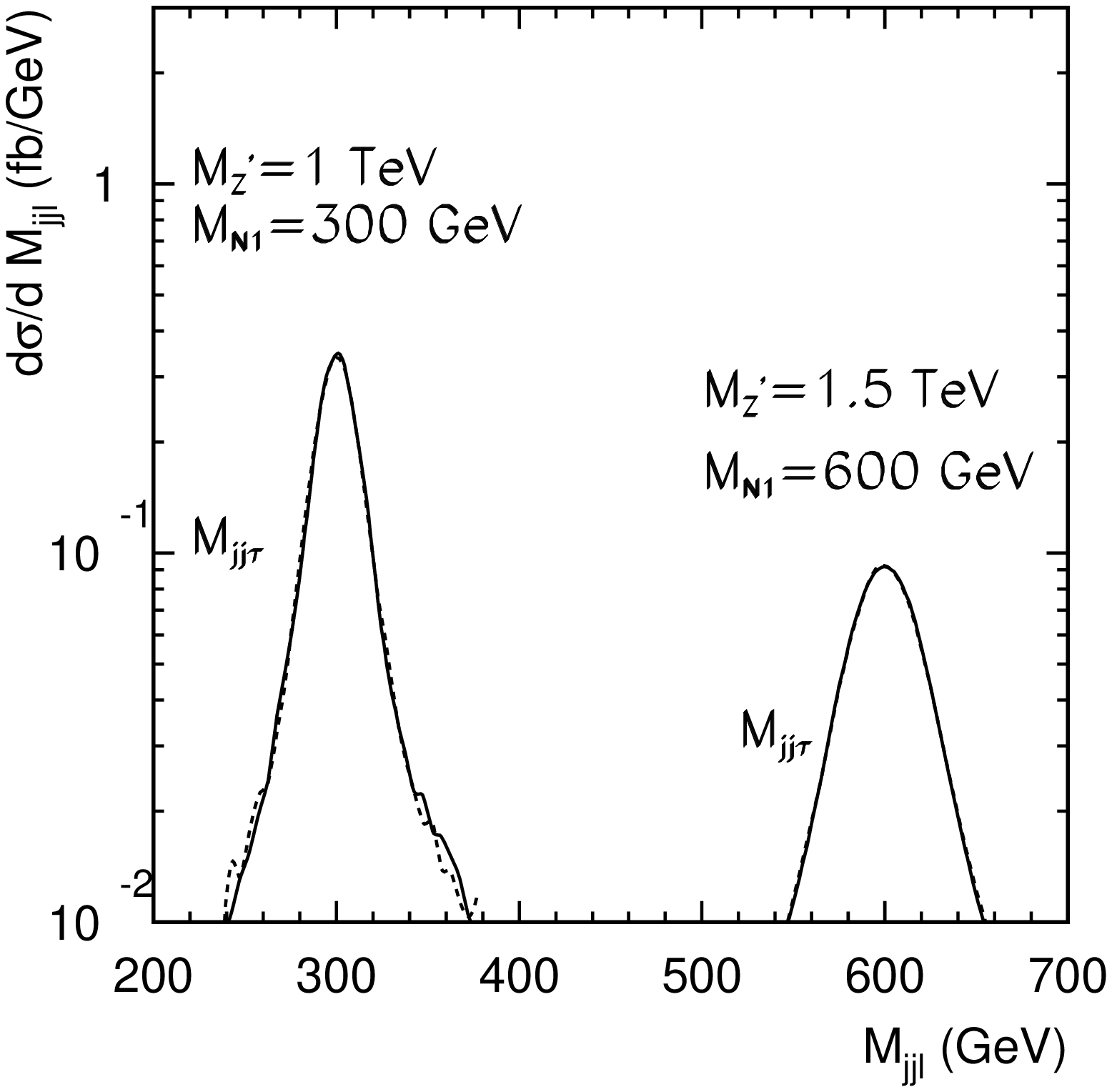}
\includegraphics[scale=1,width=8cm]{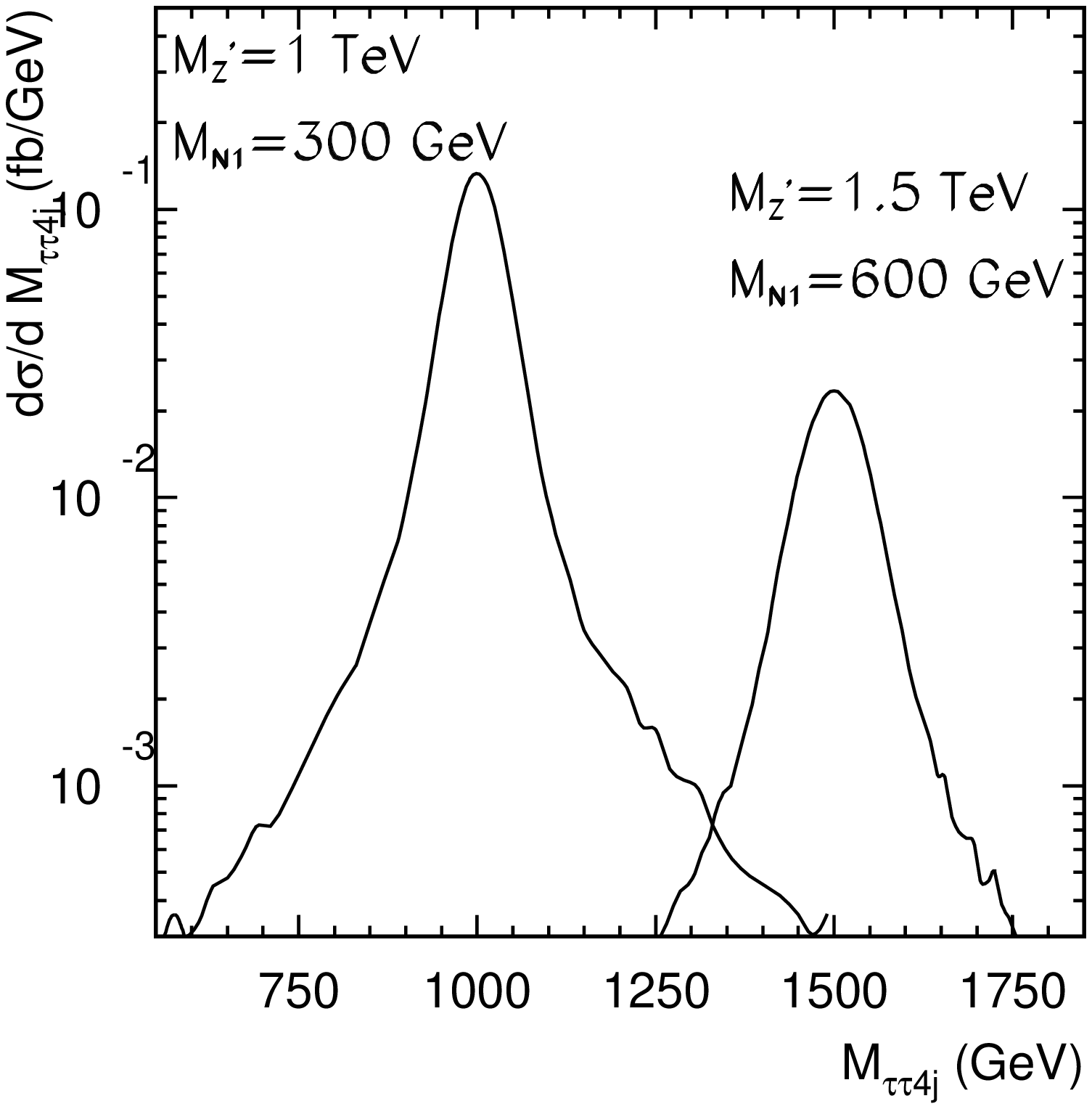}
\end{tabular}
\end{center}
\caption{Reconstructed invariant mass of $M(jj\tau)$ for
$M_{N_1}=300,\  600~{\rm GeV}$, respectively (left), and $M(\tau\tau
4j)$ for $M_{Z'}=1,\  1.5~{\rm TeV}$ (right).} \label{ren2t}
\end{figure}

For $\tau\tau jjjj$ events with two $\tau$'s, we generalize the
momenta reconstruction to
\begin{eqnarray}
\overrightarrow{p}({\rm invisible})=\kappa_1\overrightarrow{p}({\rm
track}_1)+\kappa_2\overrightarrow{p}({\rm track}_2).
\end{eqnarray}
The proportionality constants $\kappa_1,\kappa_2$ can be determined
from the missing energy measurement as long as the two charge tracks
are linearly independent.
The $N_1$ pair kinematics can be  once again fully
reconstructed. The reconstructed invariant masses of $M(\tau jj)$ is
plotted in Fig.~\ref{ren2t}.
The nice mass peaks of the $\tau jj$ system at $M_{N_1}$ and  the
$\tau\tau jj$ system at $M_{Z'}$ make the signal stand out of the SM backgrounds.

\begin{figure}[tb]
\begin{center}
\begin{tabular}{cc}
\includegraphics[scale=1,width=8.5cm]{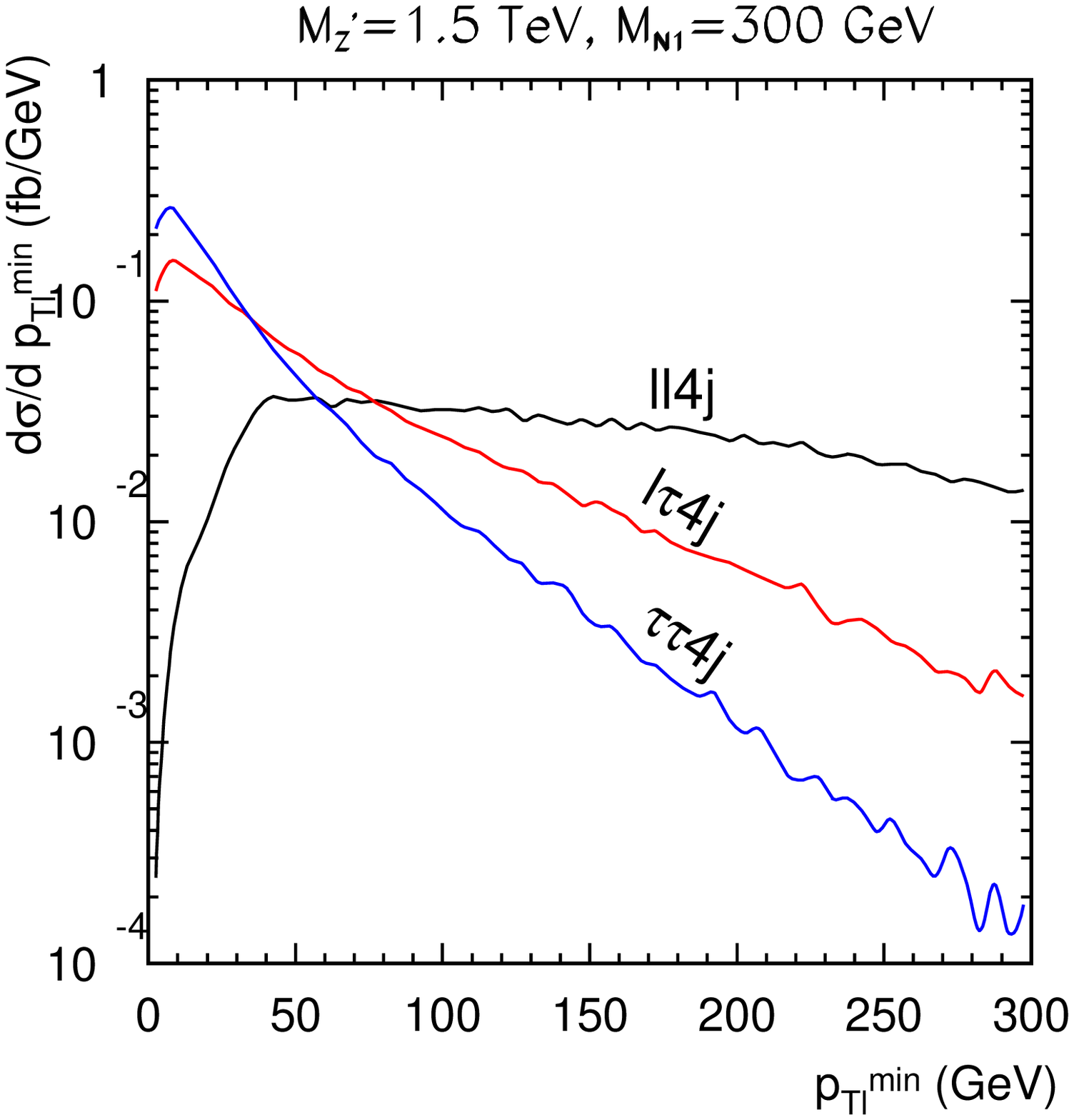}
\end{tabular}
\end{center}
\caption{$p_T$ distribution of the softer lepton from the $N_1$ and
$\tau$ decays in the events of $\ell\ell jjjj$, $\ell\tau jjjj$ and
$\tau\tau jjjj$, for a $N_1$ mass $300~{\rm GeV}$ and $Z'$ mass
$1.5~{\rm TeV}$.} \label{ptl}
\end{figure}

It is important to note a difference between the leptons from the primary
$N_1$ decay and from the $\tau$ decay: the latter is much softer.
In Fig.~\ref{ptl} we show the $p_T$ distribution of the softer
lepton from the $N_1$ and $\tau$ decays in the events of $\ell\ell
jjjj$, $\ell\tau jjjj$ and  $\tau\tau jjjj$.
This feature could provide additional discrimination power to separate
the three different leptonic channels if needed to fit the flavor structure for a
underlying theory.

\subsection{Measuring Branching Fractions and Probing the Neutrino Mass Patterns}
So far, we have only studied the characteristic features of the
signal and backgrounds for the leading channels and have not
included the proper branching fractions for the individual lepton flavors.
For illustration, consider first the cleanest channel, $N_1N_1\to
e^\pm e^\pm jjjj$. The number of events is written as
\begin{eqnarray}
N=L\times \sigma(pp\to N_1N_1)\times 2 \ {\rm BR}^2(N_1\to e^+ W^-)
({6\over 9})^2,\label{event}
\end{eqnarray}
where $L$ is the integrated luminosity and the factor (6/9) is due the the $W$
hadronic decay.
Given a sufficient number of
events $N$, the mass of $N_1$ is determined by the invariant mass of
lepton and jet $M_{\ell jj}$. We thus predict the corresponding
production rate $\sigma(pp\to N_1N_1)$ for this given mass. The only
unknown in the Eq.~(\ref{event}) is the decay branching fraction.

\begin{figure}[tb]
\begin{center}
\begin{tabular}{cc}
\includegraphics[scale=1,width=14cm]{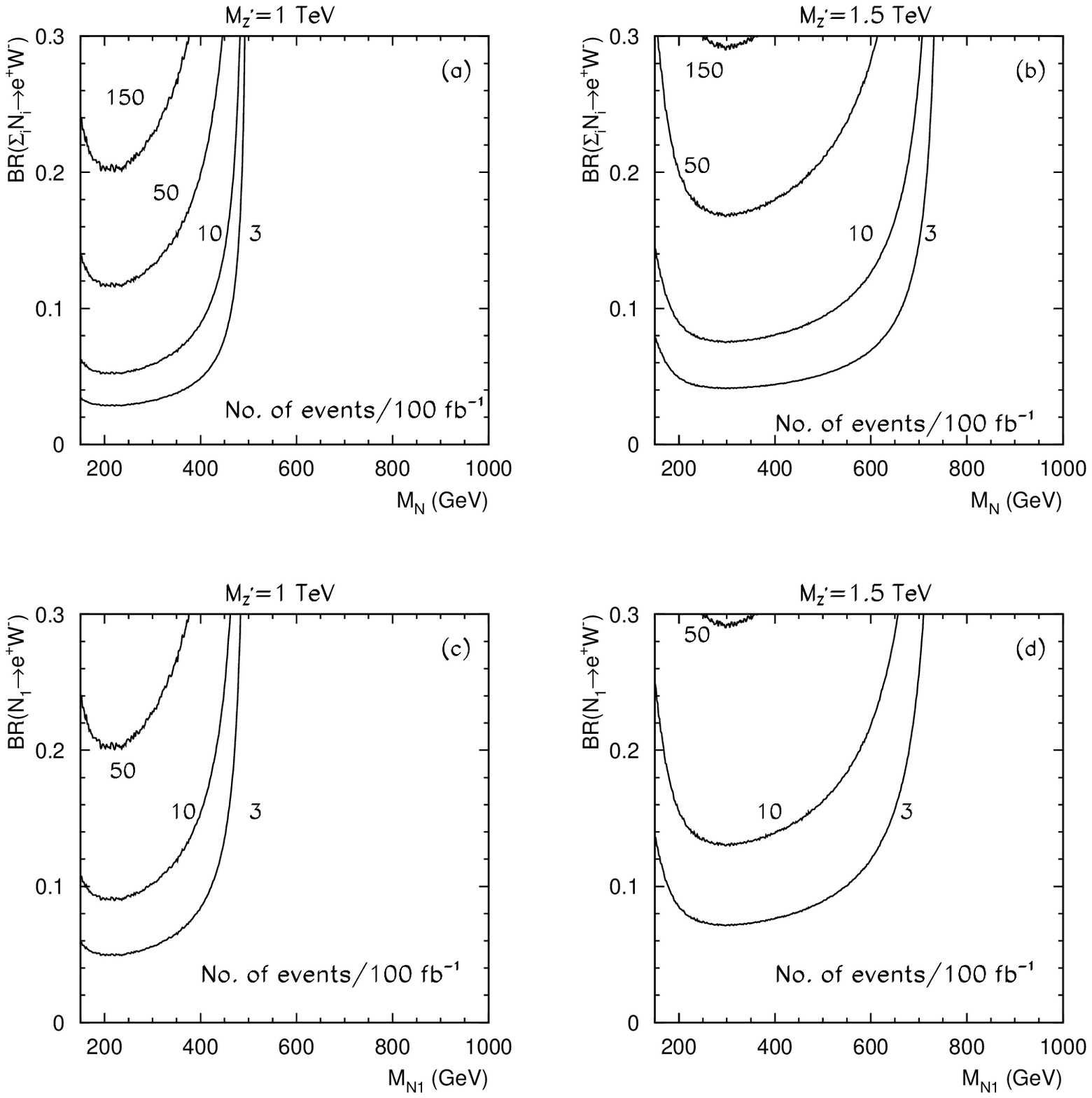}
\end{tabular}
\end{center}
\caption{Event contours in the BR$-M_{N}$ plane at the LHC with an
integrated luminosity $100~{\rm fb}^{-1}$ for degenerate case
$\sum_{i=1,2,3}N_iN_i\to e^+e^+ W^-W^-$ with (a) $M_{Z'}=1~{\rm
TeV}$, (b) $M_{Z'}=1.5~{\rm TeV}$, and for non-degenerate case
$N_1N_1\to e^+e^+ W^-W^-$ with (c) $M_{Z'}=1~{\rm TeV}$ and (d)
$M_{Z'}=1.5~{\rm TeV}$, including all the judicious cuts describe in
the early sections.} \label{eve}
\end{figure}

We present the event contours in the BR$-M_{N}$ plane in
Fig.~\ref{eve} for $100~{\rm fb}^{-1}$ luminosity and degenerate
case with (a) $M_{Z'}=1~{\rm TeV}$ and (b) $M_{Z'}=1.5~{\rm TeV}$
including all the judicious cuts described earlier, with which the
backgrounds are insignificant.

In Fig.~\ref{eve} (c) and (d), we show the event contours in the
BR$-M_{N_1}$ plane, for $100~{\rm fb}^{-1}$ luminosity and
non-degenerate case including all the judicious cuts described
earlier.
We see that the reach to a low BR can be quite encouraging.

As we presented earlier, the $N_1$ decay branching
fractions and the light neutrino mass matrix are directly
correlated. Measuring the BR's of different flavor combinations
becomes crucial in understanding the neutrino mass pattern and thus
the mass generation mechanism.
In the degenerate case, we have the prediction for the flavor combinations
\beq {\rm BR}(NN \to \ell\ell WW) \approx  \left\{
\begin{array}{ll}
\displaystyle 2 \times (23\%)^2 \quad &{\rm for\ NH:}\  (\mu^\pm+\tau^\pm) (\mu^\pm+\tau^\pm) WW,  \\ [1mm]
\displaystyle 2 \times (13\%)^2 \quad &{\rm for\ IH:}\  e^\pm e^\pm WW,  \\ [1mm]
\displaystyle 2 \times (17\%)^2 \quad &{\rm for\ QD:}\  (e^\pm+\mu^\pm+\tau^\pm) (e^\pm+\mu^\pm+\tau^\pm) WW,
\end{array}
\right. \label{BRN}
\eeq
for $\Phi_1=\Phi_2=0$, independent of the matrix $\Omega$. On the other hand, for the non-degenerate
situation,  the flavor prediction is like
\beq {\rm BR}(NN \to \ell\ell WW) \approx  \left\{
\begin{array}{ll}
\displaystyle 2 \times (20\%)^2 \quad &{\rm for\ N_1:}\ e^\pm e^\pm WW,  \\ [1mm]
\displaystyle 2 \times (17\%)^2 \quad &{\rm for\ N_2:}\
(e^\pm+\mu^\pm+\tau^\pm) (e^\pm+\mu^\pm+\tau^\pm) WW, \\ [1mm]
\displaystyle 2 \times (23\%)^2 \quad &{\rm for\  N_3:}\  (\mu^\pm+\tau^\pm) (\mu^\pm+\tau^\pm) WW,
\end{array}
\right. \label{BRNis}
\eeq
for $\Omega=I$, independent of the neutrino mass patterns as well as  $\Phi_1,\ \Phi_2$.
These predictions  are the consequence from the low energy oscillation experiments and this are
subject to test at the LHC to confirm the theory.

\begin{figure}[tb]
\begin{center}
\begin{tabular}{cc}
\includegraphics[scale=1,width=8cm]{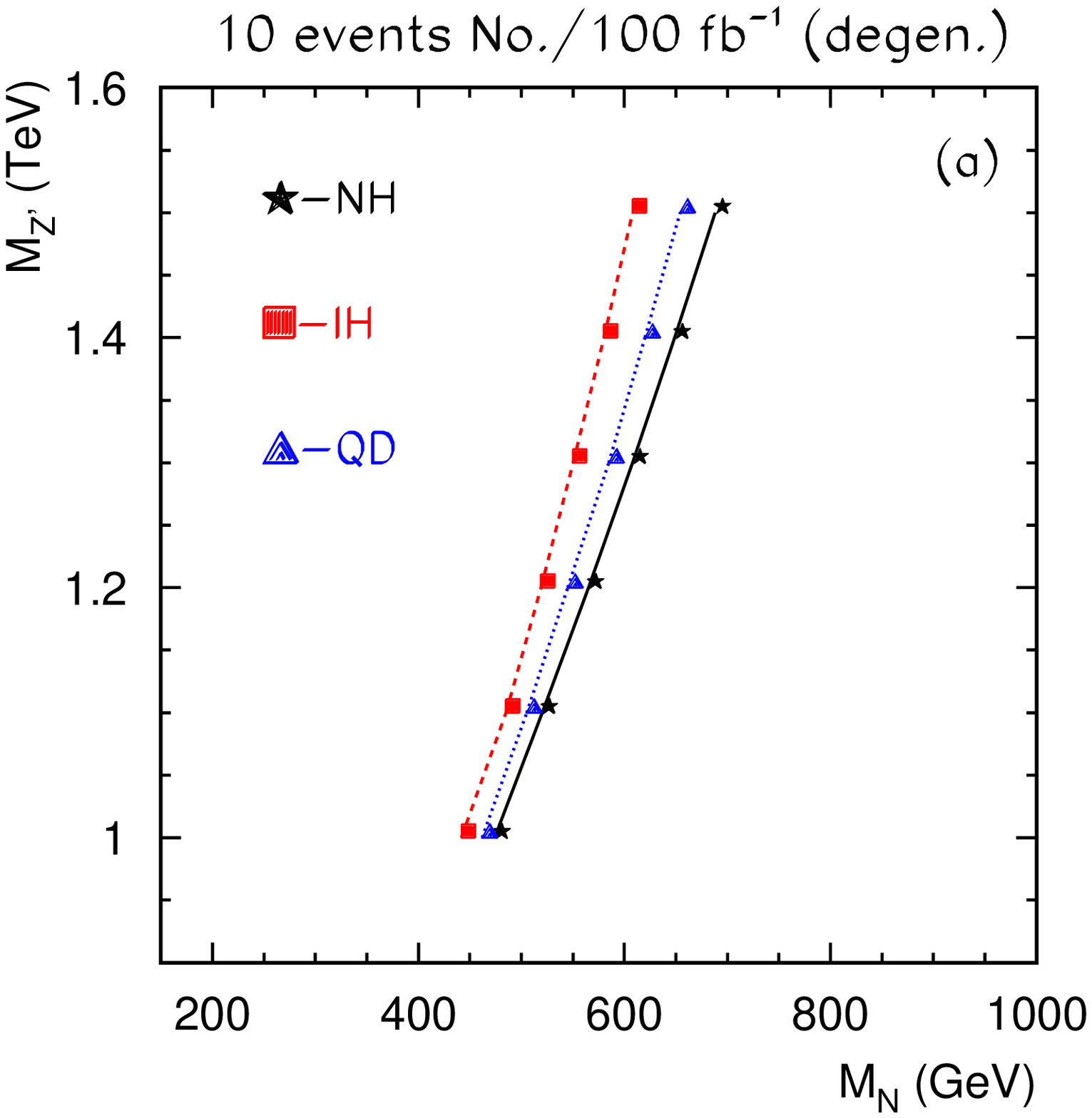}
\includegraphics[scale=1,width=8cm]{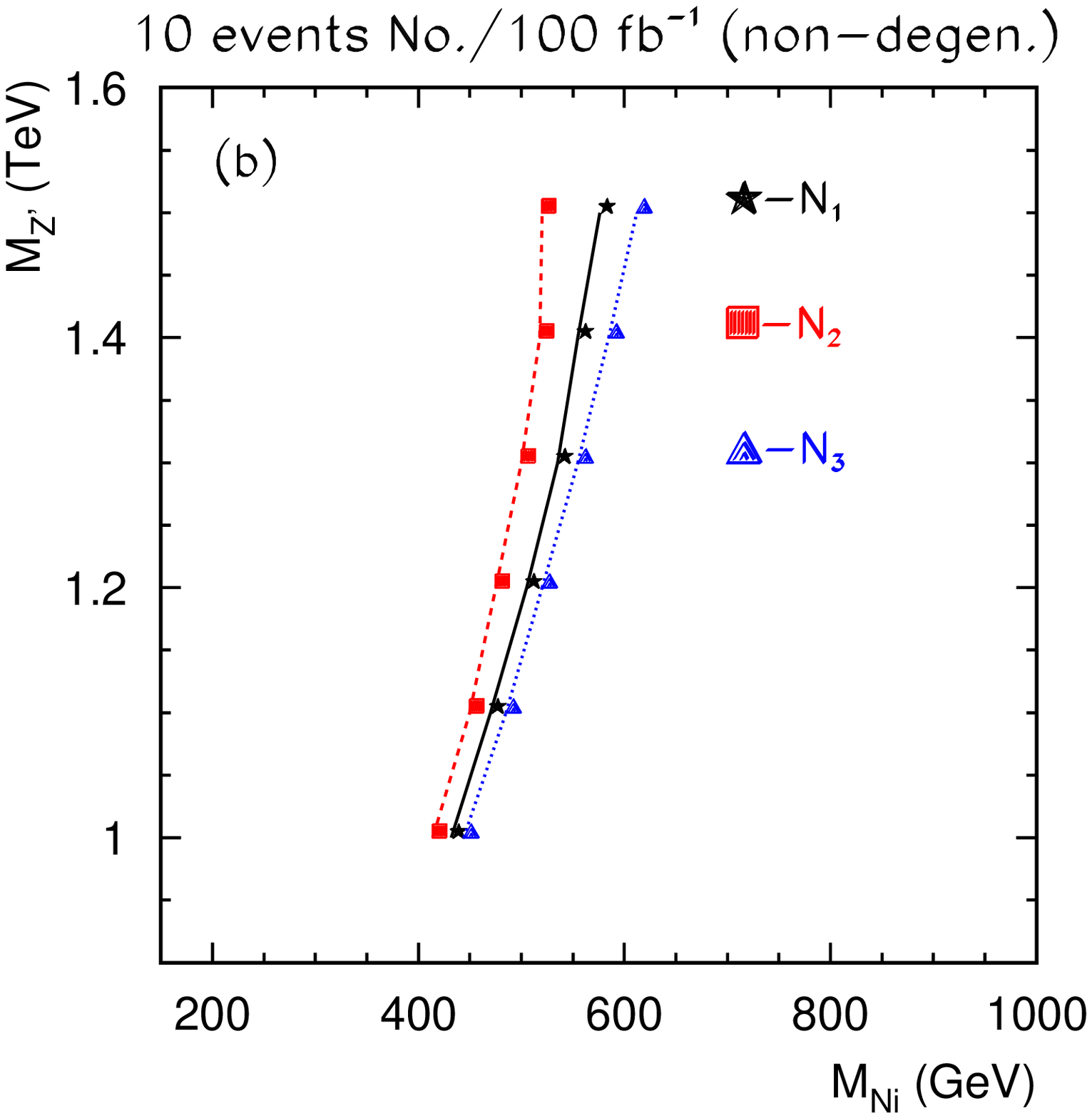}
\end{tabular}
\end{center}
\caption{Event contours in the $M_{Z'}-M_{N}$ plane at the LHC with
an integrated luminosity $100~{\rm fb}^{-1}$ and 10 events number
for (a) degenerate case $\sum_{i=1,2,3}N_iN_i\to \ell^\pm \ell^\pm
WW$ for NH, IH, and QD, and (b) non-degenerate case  for $N_1, N_2$
and $N_3$, all as predicted in Eqs.~(\ref{BRN}) and (\ref{BRNis}).}
\label{eve2}
\end{figure}

In Fig.~\ref{eve2} we show the event contours in the $M_{Z'}-M_{N}$
plane, for (a) production of $\sum_iN_i$ in NH (solid curve), IH
(dashed curve) and QD (dotted curve) for the degenerate case and (b)
production of $N_1$ (solid curve), $N_2$ (dashed curve) and $N_3$
(dotted curve) for the non-degenerate case with $100~{\rm fb}^{-1}$
luminosity, 10 events numbers and branching fractions of heavy
neutrinos predicted in Eqs.~(\ref{BRN}) and (\ref{BRNis}). The
values of $M_{Z'}$ and $M_{N}$ on the left-hand side of the curves
would give more than 10 events for $100~{\rm fb}^{-1}$ luminosity
and more accessible heavy neutrino decay branching fractions at the
LHC.

\newpage
\section{Summary}
In this article we have investigated the possibility to test the
so-called Type I seesaw mechanism for neutrino masses at the CERN
Large Hadron Collider in the context of two simple extensions of the
Standard Model where $B-L$ is part of the gauge symmetry. We have
studied in great detail the predictions of the right-handed neutrino
decays in each spectrum for neutrino masses showing the most
optimistic scenarios where one could hope to distinguish the
spectrum using the properties of the decays.

We have found the following interesting results:

\begin{itemize}

\item Working in the
context of two simple extensions of the Standard Model with a local gauge
symmetry $B-L$ or  $X=Y - \frac{5}{4} (B-L)$,
one can produce the heavy neutrinos through the $Z'$ gauge
boson  in each scenario. In both cases one has a dynamical
mechanism for the generation of heavy neutrino masses, related to
$M_{Z'}$.

\item In the case where the heavy neutrinos are degenerate, we show the possibility
to distinguish the neutrino spectrum.
The branching
fractions can differ by one order of magnitude in NH case with
$BR(\mu^\pm W^\mp),BR(\tau^\pm W^\mp)\gg BR(e^\pm W^\mp)$, and
a factor of a few in the IH spectrum with $BR(e^\pm
W^\mp)>BR(\mu^\pm W^\mp),BR(\tau^\pm W^\mp)$
when the Majorana phases are ignored.
As one expects,
all these channels are quite similar when the neutrino spectrum is
quasi-degenerate, $m_1\approx m_2\approx m_3\geq 0.05$ eV.

\item  In the case when $\Omega$ is an identity matrix or with only unity
entries generally, we find: $BR(e^\pm W^\mp)>BR(\mu^\pm W^\mp),\ BR(\tau^\pm W^\mp)$ for
$N_1$ decay, $BR(e^\pm W^\mp)\approx BR(\mu^\pm W^\mp)\approx
BR(\tau^\pm W^\mp)$ for $N_2$ and $BR(\mu^\pm W^\mp),\ BR(\tau^\pm
W^\mp)\gg BR(e^\pm W^\mp)$ for $N_3$ in both NH and IH. The
branching fractions in these cases are independent of Majorana
phases.

\item In general, the form of $\Omega$ governs heavy neutrino decay patterns.
Future tests on  the flavor combinations of SM charged leptons would reveal the
specific model structure.

\item The above-studied $\Delta L=2$  channels can take the search to $M_{Z'}\approx 2$ TeV
at the LHC. The sensitivity to the leptonic branching fractions of $N$ decay can be about $10\%$.

\item In particular, in a significant part of the parameter space of $M_N$
and the mixings, the $N$ decay could lead to distinctive signatures with secondary displaced
vertices. yielding essentially background-free signal for $N$'s.

\end{itemize}

\subsection*{Acknowledgment}
The work of P. F. P. was supported in part by the U.S. Department of
Energy contract No. DE-FG02-08ER41531 and in part by the Wisconsin
Alumni Research Foundation. The work of T. H. is supported in part
by the U.S. Department of Energy under grant No. DE-FG02-95ER40896,
and by the Wisconsin Alumni Research Foundation. T. L. would like to
thank Xiao-Gang He for helpful discussions. P. F. P. would like to
thank S. Blanchet for discussions.

\appendix
\section{NEUTRINO MASSES AND  MIXINGS}
The Type I seesaw scheme introduces right-handed neutrino states, in addition to the SM matter contents.
For definitiveness, we add three of them, $\nu_R^i\ (i=1,2,3)$, which are said to be sterile since they do not
carry any SM gauge quantum numbers.
The SM gauge invariant and renormalizable interactions to generate neutrino masses include both Dirac as
well as Majorana terms
\begin{eqnarray}
-{\cal L}_{\nu}^I &=& \overline{l}_{L} \ Y_\nu^D \ \tilde{H} \ \nu_{R}
\ + \
{1\over 2} \overline{(\nu^c)}_{L} \ M_N \ \nu_{R} + \ \text{h.c.}
\end{eqnarray}
where $Y_\nu^D,\ M_N$ are $3\times 3$ matrices in the generation space,
with $\tilde{H}=i \sigma_2 H^*$ and $H^T=\left( H^+ \ H^0 \right)$. Once $H$ gets the vacuum expectation value
$\langle H\rangle =v_0/\sqrt 2$, the neutrinos acquire Dirac masses $\md = Y_\nu^D \ v_0 / \sqrt{2}$,
\begin{eqnarray}
-{\cal L}_{\nu}^m = {1\over 2} \ \left( \overline{\nu}_{L} \ m_D \ \nu_{R} \ + \ \overline{(\nu^c)}_{L} \ m_D^T \ (\nu^c)_{R}
\ + \ \overline{(\nu^c)}_{L} \ M_{N} \ \nu_{R} \right) \ + \ \text{h.c.}
\end{eqnarray}
To diagonalize the mass matrix for neutrinos we introduce a $6\times 6$ unitary transformation
\begin{equation}
\left(
\begin{array}{c}
    \nu_L^{}    \\
    (\nu^c)_L \\
  \end{array}
\right) = \mathbb{N}\
\left(
\begin{array}{c}
    \nu_L^{}    \\
    (\nu^c)_L \\
  \end{array}
\right)_{mass} ,
\quad
\mathbb{N}=
\left(
\begin{array}{cc}
    U & V     \\
    V_C & U_C \\
  \end{array}
\right).
\end{equation}
Then,
\begin{eqnarray}
\mathbb{N}^\dagger
\left(
  \begin{array}{cc}
    0 & m_D \\
    m_{D}^{T} & M_N \\
  \end{array}
\right) \mathbb{N}^\ast &=&
\left(
  \begin{array}{cc}
    m & 0 \\
    0 & M  \\
  \end{array}
\right),
\end{eqnarray}
or explicitly,
\begin{eqnarray}
& & V_C^\dagger m_{D}^{T} U^\ast + U^\dagger m_D V_C^\ast +
V_C^\dagger M_N V_C^\ast = m,
\\
&& U_C^\dagger m_{D}^{T} V^\ast + V^\dagger m_D U_C^\ast +
U_C^\dagger M_N U_C^\ast = M,
\\
&& V_C^\dagger m_{D}^{T} V^\ast + U^\dagger m_D U_C^\ast +
V_C^\dagger M_N U_C^\ast = 0, \label{seesaw1}
\end{eqnarray}
where $m=diag (m_1, m_2, m_3)$ and $M=diag(M_1, M_2, M_3)$ are
diagonal matrices of the mass eigenvalues. In the limit $M_N \gg  m_D$ we
have
\beq
m \approx {m_D^2\over M_N},\quad M \approx M_N.
\eeq
Then we have three light neutrinos and three heavy ones, all Majorana-type.
Note that $U$ and $V$ mix the light and heavy neutrinos, respectively, into
the active weak interaction eigenstates of neutrinos. The mixing elements are
typically like
\beq
U^2\approx {\cal O}(1),\quad V^2\approx {m\over
M},
\eeq
and the unitarity conditions read
\begin{eqnarray}
&& U U^\dagger + V V^\dagger = U^\dagger U + V_C^\dagger V_C = V_C
V_C^\dagger + U_C U_C^\dagger =
V^\dagger V + U_C^\dagger U_C = I, \\
&&  U V_C^\dagger + V U_C^\dagger =  U^\dagger V + V_C^\dagger
U_C=0.
\end{eqnarray}
It can be shown  that the following relations hold
\begin{eqnarray}
V_C^\dagger m_{D}^{T} - m U^T = 0, \quad m_{D} U_C^\ast - V M = 0.
\end{eqnarray}
Assuming that a $3\times 3$ matrix $E$ diagonalizes the mass matrix of
the charged leptons,  we then define
\bea
&& E^\dagger U \equiv V_{PMNS},\quad E^\dagger V \equiv V_{\ell N} , \\
&&  V_{PMNS} V^\dagger_{PMNS} + V_{\ell N}V_{\ell N}^\dagger = I ,
\eea
where $V_{PMNS}$ and $V_{\ell N}$ describe the transitions between the light neutrino and
heavy neutrino to the charged leptons, respectively,  via the weak charged currents.
Note that the identification of $V_{PMNS}$ to the PMNS matrix is only approximate.
We then obtain an important relation among the physical quantities
\bea
\label{VLN}
&& V_{\ell N}^\ast \ M \ V_{\ell N}^\dagger =
-V_{PMNS}^\ast \ m \ V^\dagger_{PMNS}.
\eea
Although the masses and mixings of the light neutrinos on the right-handed side can be
measured from the oscillation experiments, it is quite involved to
solve for $V_{\ell N}$ via this set of quadratic equations.
Absorbing the minus sign on the right side of the above
equation in the definition of $V_{\ell N}$ one can write down
a formal solution with the help of an auxiliary matrix
\begin{eqnarray}
V_{\ell N}&=& \ V_{PMNS} \ m^{1/2} \ \Omega \ M^{-1/2},
\end{eqnarray}
where $\Omega$ is an orthogonal complex matrix which can be parameterized as
\begin{equation}
\Omega (w_{21}, w_{31}, w_{32})= R_{12} (w_{21})\ R_{13}(w_{31})\ R_{23}(w_{32}),
\end{equation}
with
\small{
\begin{equation}
R_{12}=\left(
\begin{array}{ccc}
    u_{21} & - w_{21}  & 0
    \\
    w_{21} &  u_{21}& 0
    \\
    0 & 0 & 1
  \end{array}
\right), \quad
%
R_{13}=\left(
\begin{array}{ccc}
    u_{31} &  0 & -w_{31}
    \\
    0 & 1 & 0
    \\
    w_{31} & 0 & u_{31}
  \end{array}
\right), \quad
%
R_{23}=\left(
\begin{array}{ccc}
    1 & 0 & 0 \\
    0 & u_{32} & - w_{32}
    \\
    0 & w_{32} & u_{32}
  \end{array}
\right).
\end{equation} }
where $u_{ij} = \pm\sqrt{1-w_{ij}^2 }$  and $-1 \leq w_{ij} \leq 1$
when the matrix $\Omega$ is real.

\section{EXPLICIT EXPRESSIONS FOR THE MIXINGS $V_{\ell N}$}

\subsection{Case I: Degenerate Heavy Neutrinos}

From Eq.~(\ref{VLN}), assuming degenerate heavy neutrinos, we have
\begin{eqnarray}
M_N\sum_N(V_{eN}^\ast)^2&=&c_{13}^2s_{12}^2m_2+c_{12}^2c_{13}^2e^{-i\Phi_1}m_1+s_{13}^2e^{i(2\delta-\Phi_2)}m_3,\\
M_N\sum_N(V_{\mu
N}^\ast)^2&=&(c_{12}c_{23}-s_{12}s_{13}s_{23}e^{-i\delta})^2m_2+(c_{23}s_{12}+
c_{12}s_{13}s_{23}e^{-i\delta})^2e^{-i\Phi_1}m_1 \nonumber \\
&+&c_{13}^2s_{23}^2e^{-i\Phi_2}m_3,\\
M_N\sum_N(V_{\tau
N}^\ast)^2&=&(c_{12}s_{23}+c_{23}s_{12}s_{13}e^{-i\delta})^2m_2+(s_{12}s_{23}-
c_{12}c_{23}s_{13}e^{-i\delta})^2e^{-i\Phi_1}m_1\nonumber \\
&+&c_{13}^2c_{23}^2e^{-i\Phi_2}m_3.
\end{eqnarray}

\subsection{Case II: Non-degenerate Heavy Neutrinos}
The general expressions for the mixing between the charged leptons and heavy neutrinos,
in terms of the neutrino oscillation parameters and the unknown matrix $\Omega$, are given by
\begin{eqnarray}
V_{e1}\sqrt{M_1}&=&\sqrt{m_2}c_{13}s_{12}w_{21}\sqrt{1-w_{31}^2}
+\sqrt{m_1}c_{12}c_{13}\sqrt{(1-w_{21}^2)(1-w_{31}^2)}e^{i\Phi_1/2}
+\sqrt{m_3}s_{13}w_{31}e^{i(\Phi_2/2-\delta)}, \nonumber
\\
\\
V_{\mu 1}\sqrt{M_1}&=&\sqrt{m_2}(c_{12}c_{23}-s_{12}s_{13}s_{23}e^{i
\delta})w_{21}\sqrt{1-w_{31}^2}
\nonumber \\
&+&\sqrt{m_1}(-s_{12}c_{23}-c_{12}s_{13}s_{23}e^{i
\delta})\sqrt{(1-w_{21}^2)(1-w_{31}^2)}e^{i\Phi_1/2}
+\sqrt{m_3}c_{13}s_{23}w_{31}e^{i\Phi_2/2},
\\
V_{\tau
1}\sqrt{M_1}&=&\sqrt{m_2}(-c_{12}s_{23}-s_{12}s_{13}c_{23}e^{i
\delta})w_{21}\sqrt{1-w_{31}^2}
\nonumber \\
&+&\sqrt{m_1}(s_{12}s_{23}-c_{12}s_{13}c_{23}e^{i
\delta})\sqrt{(1-w_{21}^2)(1-w_{31}^2)}e^{i\Phi_1/2}
+\sqrt{m_3}c_{13}c_{23}w_{31}e^{i\Phi_2/2}.
\end{eqnarray}

\begin{eqnarray}
V_{e2}\sqrt{M_2}&=&\sqrt{m_2}c_{13}s_{12}(-w_{21}w_{31}w_{32}+\sqrt{(1-w_{21}^2)(1-w_{32}^2)})\nonumber
\\
&+&\sqrt{m_1}c_{12}c_{13}(-w_{31}w_{32}\sqrt{1-w_{21}^2}-w_{21}\sqrt{1-w_{32}^2})e^{i\Phi_1/2}
+\sqrt{m_3}s_{13}w_{32}\sqrt{1-w_{31}^2}e^{i(\Phi_2/2-\delta)},
\nonumber
\\
\\
V_{\mu
2}\sqrt{M_2}&=&\sqrt{m_2}(c_{12}c_{23}-s_{12}s_{13}s_{23}e^{i\delta})
(-w_{21}w_{31}w_{32}+\sqrt{(1-w_{21}^2)(1-w_{32}^2)})\nonumber
\\
&+&\sqrt{m_1}(-s_{12}c_{23}-c_{12}s_{13}s_{23}e^{i\delta})
(-w_{32}w_{31}\sqrt{1-w_{21}^2}-w_{21}\sqrt{1-w_{32}^2})e^{i\Phi_1/2}\nonumber
\\
&+&\sqrt{m_3}c_{13}s_{23}w_{32}\sqrt{1-w_{31}^2}e^{i\Phi_2/2},
\\
V_{\tau
2}\sqrt{M_2}&=&\sqrt{m_2}(-c_{12}s_{23}-s_{12}s_{13}c_{23}e^{i\delta})
(-w_{21}w_{31}w_{32}+\sqrt{(1-w_{21}^2)(1-w_{32}^2)})\nonumber
\\
&+&\sqrt{m_1}(s_{12}s_{23}-c_{12}s_{13}c_{23}e^{i\delta})
(-w_{32}w_{31}\sqrt{1-w_{21}^2}-w_{21}\sqrt{1-w_{32}^2})e^{i\Phi_1/2}\nonumber
\\
&+&\sqrt{m_3}c_{13}c_{23}w_{32}\sqrt{1-w_{31}^2}e^{i\Phi_2/2}.
\end{eqnarray}

\begin{eqnarray}
V_{e3}\sqrt{M_3}&=&\sqrt{m_2}c_{13}s_{12}(-w_{32}\sqrt{1-w_{21}^2}-w_{21}w_{31}\sqrt{1-w_{32}^2})\nonumber
\\
&+&\sqrt{m_1}c_{12}c_{13}(w_{21}w_{32}-w_{31}\sqrt{(1-w_{21}^2)(1-w_{32}^2)})e^{i\Phi_1/2}
\nonumber
\\
&+& \sqrt{m_3}s_{13}\sqrt{(1-w_{31}^2)(1-w_{32}^2)}e^{i(\Phi_2/2-\delta)},
\nonumber
\\
\\
V_{\mu
3}\sqrt{M_3}&=&\sqrt{m_2}(c_{12}c_{23}-s_{12}s_{13}s_{23}e^{i\delta})
(-w_{32}\sqrt{1-w_{21}^2}-w_{21}w_{31}\sqrt{1-w_{32}^2})\nonumber
\\
&+&\sqrt{m_1}(-s_{12}c_{23}-c_{12}s_{13}s_{23}e^{i\delta})
(w_{32}w_{21}-w_{31}\sqrt{(1-w_{21}^2)(1-w_{32}^2)})e^{i\Phi_1/2}\nonumber
\\
&+&\sqrt{m_3}c_{13}s_{23}\sqrt{(1-w_{31}^2)(1-w_{32}^2)}e^{i\Phi_2/2},
\\
V_{\tau
3}\sqrt{M_3}&=&\sqrt{m_2}(-c_{12}s_{23}-s_{12}s_{13}c_{23}e^{i\delta})
(-w_{32}\sqrt{1-w_{21}^2}-w_{21}w_{31}\sqrt{1-w_{32}^2})\nonumber
\\
&+&\sqrt{m_1}(s_{12}s_{23}-c_{12}s_{13}c_{23}e^{i\delta})
(w_{32}w_{21}-w_{31}\sqrt{(1-w_{21}^2)(1-w_{32}^2)})e^{i\Phi_1/2}\nonumber
\\
&+&\sqrt{m_3}c_{13}c_{23}\sqrt{(1-w_{31}^2)(1-w_{32}^2)}e^{i\Phi_2/2}.
\end{eqnarray}

We now present the two cases according to the light neutrino mass
spectra, assuming $m_{1(3)}\approx 0$ and $s_{13}=0$.

\begin{itemize}
\item Normal Hierarchy:\\
Under the good approximations $m_1\approx 0$ and $s_{13}=0$,
one finds the following expressions
\begin{eqnarray}
M_1|V_{e1}|^2&\approx&\sqrt{\Delta
m_{21}^2}(s_{12}w_{21}\sqrt{1-w_{31}^2})^2,
\\
M_1|V_{\mu 1}|^2&\approx&|\sqrt[4]{\Delta
m_{21}^2}c_{12}c_{23}w_{21}\sqrt{1-w_{31}^2}+\sqrt[4]{\Delta
m_{31}^2}s_{23}w_{31}e^{i\Phi_2/2}|^2,
\\
M_1|V_{\tau 1}|^2&\approx&|\sqrt[4]{\Delta
m_{21}^2}c_{12}s_{23}w_{21}\sqrt{1-w_{31}^2}-\sqrt[4]{\Delta
m_{31}^2}s_{23}w_{31}e^{i\Phi_2/2}|^2.
\end{eqnarray}
\begin{eqnarray}
M_2|V_{e2}|^2&\approx&\sqrt{\Delta
m_{21}^2}s_{12}^2(-w_{21}w_{31}w_{32}+\sqrt{(1-w_{21}^2)(1-w_{32}^2)})^2,
\\
M_2|V_{\mu 2}|^2&\approx&|\sqrt[4]{\Delta
m_{21}^2}c_{12}c_{23}(-w_{21}w_{31}w_{32}+\sqrt{(1-w_{21}^2)(1-w_{32}^2)})
\nonumber
\\
&+& \sqrt[4]{\Delta
m_{31}^2}s_{23}w_{32}\sqrt{1-w_{31}^2}e^{i\Phi_2/2}|^2,
\\
M_2|V_{\tau 2}|^2&\approx&|\sqrt[4]{\Delta
m_{21}^2}c_{12}s_{23}(-w_{21}w_{31}w_{32}+\sqrt{(1-w_{21}^2)(1-w_{32}^2)})
\nonumber \\
&-& \sqrt[4]{\Delta
m_{31}^2}c_{23}w_{32}\sqrt{1-w_{31}^2}e^{i\Phi_2/2}|^2.
\end{eqnarray}
\begin{eqnarray}
M_3|V_{e3}|^2&\approx&\sqrt{\Delta
m_{21}^2}s_{12}^2(-w_{32}\sqrt{1-w_{21}^2}-w_{21}w_{31}\sqrt{1-w_{32}^2})^2,
\\
M_3|V_{\mu 3}|^2&\approx&|\sqrt[4]{\Delta
m_{21}^2}c_{12}c_{23}(-w_{32}\sqrt{1-w_{21}^2}-w_{21}w_{31}\sqrt{1-w_{32}^2})
\nonumber \\
&+ & \sqrt[4]{\Delta
m_{31}^2}s_{23}\sqrt{(1-w_{31}^2)(1-w_{32}^2)}e^{i\Phi_2/2}|^2,
\\
M_3|V_{\tau 3}|^2&\approx&|\sqrt[4]{\Delta
m_{21}^2}c_{12}s_{23}(-w_{32}\sqrt{1-w_{21}^2}-w_{21}w_{31}\sqrt{1-w_{32}^2})
\nonumber \\
&-& \sqrt[4]{\Delta
m_{31}^2}c_{23}\sqrt{(1-w_{31}^2)(1-w_{32}^2)}e^{i\Phi_2/2}|^2.
\end{eqnarray}
\item Inverted Hierarchy:\\
Under the approximations $m_3\approx 0$ and $s_{13}=0$, we have
\begin{eqnarray}
M_1|V_{e1}|^2&\approx&|\sqrt[4]{\Delta m_{21}^2+|\Delta
m_{31}^2|}s_{12}w_{21}\sqrt{1-w_{31}^2}
\nonumber \\
&+& \sqrt[4]{|\Delta
m_{31}^2|}c_{12}\sqrt{(1-w_{31}^2)(1-w_{21}^2)}e^{i\Phi_1/2}|^2,
\\
M_1|V_{\mu 1}|^2&\approx&|\sqrt[4]{\Delta m_{21}^2+|\Delta
m_{31}^2|}c_{12}c_{23}w_{21}\sqrt{1-w_{31}^2}
\nonumber \\
&-& \sqrt[4]{|\Delta
m_{31}^2|}s_{12}c_{23}\sqrt{(1-w_{21}^2)(1-w_{31}^2)}e^{i\Phi_1/2}|^2,
\\
M_1|V_{\tau 1}|^2&\approx&|\sqrt[4]{\Delta m_{21}^2+|\Delta
m_{31}^2|}c_{12}s_{23}w_{21}\sqrt{1-w_{31}^2}
\nonumber \\
&-& \sqrt[4]{|\Delta
m_{31}^2|}s_{12}s_{23}\sqrt{(1-w_{21}^2)(1-w_{31}^2)}e^{i\Phi_1/2}|^2.
\end{eqnarray}
\begin{eqnarray}
M_2|V_{e2}|^2&\approx&|\sqrt[4]{\Delta m_{21}^2+|\Delta
m_{31}^2|}s_{12}(-w_{21}w_{31}w_{32}+\sqrt{(1-w_{21}^2)(1-w_{32}^2)})\nonumber
\nonumber \\
&+&\sqrt[4]{|\Delta
m_{31}^2|}c_{12}(-w_{31}w_{32}\sqrt{1-w_{21}^2}-w_{21}\sqrt{1-w_{32}^2})e^{i\Phi_1/2}|^2,
\\
M_2|V_{\mu 2}|^2&\approx&|\sqrt[4]{\Delta m_{21}^2+|\Delta
m_{31}^2|}c_{12}c_{23}(-w_{21}w_{31}w_{32}+\sqrt{(1-w_{21}^2)(1-w_{32}^2)})
\nonumber
\\
&-&\sqrt[4]{|\Delta
m_{31}^2|}s_{12}c_{23}(-w_{31}w_{32}\sqrt{1-w_{21}^2}-w_{21}\sqrt{1-w_{32}^2})e^{i\Phi_1/2}|^2,
\\
M_2|V_{\tau 2}|^2&\approx&|\sqrt[4]{\Delta m_{21}^2+|\Delta
m_{31}^2|}c_{12}s_{23}(-w_{21}w_{31}w_{32}+\sqrt{(1-w_{21}^2)(1-w_{32}^2)})
\nonumber
\\
&-&\sqrt[4]{|\Delta
m_{31}^2|}s_{12}s_{23}(-w_{31}w_{32}\sqrt{1-w_{21}^2}-w_{21}\sqrt{1-w_{32}^2})e^{i\Phi_1/2}|^2.
\end{eqnarray}
\begin{eqnarray}
M_3|V_{e3}|^2&\approx&|\sqrt[4]{\Delta m_{21}^2+|\Delta
m_{31}^2|}s_{12}(-w_{32}\sqrt{1-w_{21}^2}-w_{21}w_{31}\sqrt{1-w_{32}^2})
\nonumber
\\
&+&\sqrt[4]{|\Delta
m_{31}^2|}c_{12}(w_{21}w_{32}-w_{31}\sqrt{(1-w_{21}^2)(1-w_{32}^2)})e^{i\Phi_1/2}|^2,
\\
M_3|V_{\mu 3}|^2&\approx&|\sqrt[4]{\Delta m_{21}^2+|\Delta
m_{31}^2|}c_{12}c_{23}(-w_{32}\sqrt{1-w_{21}^2}-w_{21}w_{31}\sqrt{1-w_{32}^2})
\nonumber
\\
&-&\sqrt[4]{|\Delta
m_{31}^2|}s_{12}c_{23}(w_{21}w_{32}-w_{31}\sqrt{(1-w_{21}^2)(1-w_{32}^2)})e^{i\Phi_1/2}|^2,
\\
M_3|V_{\tau 3}|^2&\approx&|\sqrt[4]{\Delta m_{21}^2+|\Delta
m_{31}^2|}c_{12}s_{23}(-w_{32}\sqrt{1-w_{21}^2}-w_{21}w_{31}\sqrt{1-w_{32}^2})
\nonumber
\\
&-&\sqrt[4]{|\Delta
m_{31}^2|}s_{12}s_{23}(w_{21}w_{32}-w_{31}\sqrt{(1-w_{21}^2)(1-w_{32}^2)})e^{i\Phi_1/2}|^2.
\end{eqnarray}
\end{itemize}
%

\section{$U(1)_{B-L}$ AND $U(1)_{X}$ EXTENSIONS OF THE STANDARD MODEL}
It is well-known that $B-L$ is an accidental global
symmetry in the Standard Model and its origin is unknown.
In order to understand the origin of Majorana neutrino masses
it is crucial to look for new scenarios where $B-L$ can be
spontaneously broken. Here we focus on a simple extension of
the Standard Model where $U(1)_{B-L}$ is a local symmetry
and in order to cancel the anomalies one has to introduce
three right-handed neutrinos. Therefore, this model is based
on the gauge symmetry $SU(3)_C \bigoplus SU(2)_L \bigoplus
U(1)_Y \bigoplus U(1)_{B-L}$~\cite{B-L}. The matter fields have the
following properties:
\begin{equation}
Q_L = \left(
\begin{array} {c}
u \\ d
\end{array}
\right)_L \ \sim \ (3,2,1/6,1/3), \ u_R \sim (3,1,2/3,1/3), \ d_R \
\sim \ (3,1,-1/3,1/3),
\end{equation}
\begin{equation}
l_L = \left(
\begin{array} {c}
 \nu \\ e
\end{array}
\right)_L \ \sim \ (1,2,-1/2,-1), \ e_R \sim (1,1,-1,-1), \
\text{and} \ \nu_R \sim (1,1,0,-1),
\end{equation}
where $\nu_R$ are the right-handed neutrinos. Here we use the
normalization where $Q=T_{3} + Y$. In order to generate the
right-handed neutrino masses and break the local $B-L$ symmetry one
has to add a new scalar field $S \sim (1,1,0,2)$.
\subsection{Interactions and Symmetry Breaking}
In this context the Kinetic terms for the Abelian sector are given
by
\begin{equation}
{\cal L}_{gauge}= - \frac{1}{4} F^{\mu \nu} F_{\mu \nu} \ - \
\frac{1}{4} F'^{\mu \nu} F_{\mu \nu}' - \frac{\epsilon}{2} F^{\mu
\nu} F_{\mu \nu}', \label{Kinetic}
\end{equation}
where
\begin{equation}
F^{\mu \nu}=\partial^\mu B^\nu \ - \ \partial^\nu B^\mu, \
\text{and} \ \ F_{\mu \nu}'= \partial_\mu B_\nu' \ - \
\partial_\nu B_\mu'.
\end{equation}
Here $B_\nu$ and $B_\nu'$ are the gauge fields for $U(1)_Y$ and
$U(1)_{B-L}$, respectively. Since the mixing between the Abelian
gauge bosons have to be very small we work in the case where
$\epsilon=0$. The Kinetic terms for the matter fields read as:
\begin{eqnarray}
{\cal L}_{Kinetic} &=& i \overline{Q}_L \gamma^\mu D_\mu Q_L \ + \ i
\bar{u}_R \gamma^\mu D_\mu u_R \ + \ i \bar{d}_R \gamma^\mu D_\mu
d_R
\nonumber \\
&+& i \bar{l}_L \gamma^\mu D_\mu l_L \ + \ i \bar{e}_R \gamma^\mu D_\mu e_R \ + \ i \bar{\nu}_R \gamma^\mu D_\mu \nu_R,
\end{eqnarray}
where
\begin{equation}
D_\mu \nu_R = \partial_\mu \nu_R \ - i g_{BL} B_\mu' \nu_R.
\end{equation}
As we have explained before the Higgs sector is composed of the SM
Higgs, $H^T = (H^+, H^0)$, and an extra Higgs, $S= S_R \ + \ i S_I$,
which is needed to break $B-L$. The relevant Lagrangian for the
scalar fields is given by
\begin{eqnarray}
{\cal L}_{Higgs} &=& \left( D_\mu H \right)^\dagger \left( D^\mu H
\right) \ + \ \left( D_\mu S \right)^\dagger \left( D^\mu S \right)
\ - \ V (H,S) \label{higgskin}
\end{eqnarray}
where
\begin{equation}
D_\mu S = \partial_\mu S \ + \ i 2 g_{BL} B_\mu' S.
\end{equation}
The gauge invariant Yukawa interactions of neutrinos are
\begin{eqnarray}
- {\cal L}^{\nu}_Y  &  =  & Y_\nu^D \ \bar{l}_L \ \tilde{H} \ \nu_R
\ + \ \frac{Y_\nu^M}{2} \ \nu_R^T \ C \ \nu_R \ S+ \ \rm{h.c.} .
\label{Y}
\end{eqnarray}
Once $S$ gets the vacuum expectation value $\langle S\rangle \to
v_S/\sqrt{2}$, $B-L$ is broken and one gets the mass of neutral
gauge boson $Z'=Z_{B-L}$ with $M_{Z'}=2g_{BL}v_S$ from the second
kinetic term in Eq.~(\ref{higgskin}), and the mass matrix of
right-handed neutrino with $M_N=Y_\nu^M v_S/\sqrt{2}$ from
Eq.~(\ref{Y}). In order not to upset applicability of perturbative
theory, we require $Y_\nu^M\leq 1$ and get an upper bound of the
mass of the heavy neutrinos $M_N\leq M_{Z'}/(2\sqrt{2}g_{BL})$.

The scalar potential is given by
\begin{equation}
V(H,S) = - m_H^2 \ H^\dagger H \ + \ \lambda_H \left( H^\dagger H \right)^2 \ - \ m_{S}^2 \ S^\dagger S
\ + \ \lambda_{S} \left( S^\dagger S \right)^2 \ + \ a_S \left( H^\dagger H \right) \left( S^\dagger S \right)
\end{equation}
where all parameters are real. Notice that this scalar potential has
the global symmetry $O(4)_H \bigotimes O(2)_S$. The minimization
conditions in this case read as
\begin{eqnarray}
0 &=& v_0 \left( -m_H^2 \ + \ \lambda_H v_0^2 \ + \ \frac{a_S}{2} v_S^2\right),
\\
0 &=& v_S \left( -m_S^2 \ + \ \lambda_S v_S^2 \ + \ \frac{a_S}{2} v_0^2\right).
\label{minimization}
\end{eqnarray}
Notice that one can have several vacua but only the case $v_0 \neq 0$ and $v_S \neq 0$
is allowed by the experiment. Now, in order to satisfy the condition of minimum one
has to satisfy the following condition:
\begin{equation}
\lambda_H a_S v_0^4 \ + \ 4 \lambda_H \lambda_S v_0^2 v_S^2 \ + \ \lambda_S a_S v_S^4 \ > \ 0.
\end{equation}
The potential is bounded from below when $\lambda_H \lambda_S - a_S^2/4 > 0$.
Using the minimization conditions above one can find the solution in the
phenomenological allowed case:
\begin{eqnarray}
v_0^2 = \frac{2 \left( a_S m_S^2 - 2 \lambda_S m_H^2 \right)}{ a_S^2 - 4 \lambda_H \lambda_S} > 0,
\\
v_S^2 = \frac{2 \left( a_S m_H^2 - 2 \lambda_H m_S^2 \right)}{ a_S^2 - 4 \lambda_H \lambda_S} > 0.
\end{eqnarray}
Using these conditions one can discuss different cases for the parameters in the
Lagrangian. Expressing the numerators and the denominator as
$n_1 = a_S \ m_H^2 - 2 \ \lambda_H \ m_S^2$, $n_2=a_S \ m_S^2 - 2 \ \lambda_S \ m_H^2$
and $d=a_S^2 - 4 \ \lambda_S \ \lambda_H$.

\begin{itemize}

\item Imposing $v_S^2 > 0$ one has the case $n_1 > 0$ and $d > 0$, or $n_1 < 0$ and $d < 0$.

\item Imposing $v_0^2 > 0$ one has $n_2 > 0$ and $d > 0$, or $n_2 < 0$ and $d< 0$.

\end{itemize}

\subsection{Higgs Bosons Properties}
As we have discussed before the Higgs sector of this model is
composed of the SM Higgs, $H^T = (H^+, ( v_0 + H^0 + i \xi^0
)/\sqrt{2})$, and an extra Higgs, $S= ( v_S \ + \ S^0 \ + \ i S_I )
/ \sqrt{2}$, which is needed to break $B-L$ and generate neutrino
masses. In this context one will have only two CP-even physical
Higgses $h$ and $H$, and the mass matrix for the these fields is
given by
\begin{equation}
\label{eq:massmtrx}
{\cal M}_{0}^2 =
\left( \begin{array} {cc}
\lambda_H v_0^2/2 - a_S v_S^2/4 &  a_S v_0 v_S
\\
a_S v_0 v_S  & \lambda_S v_S^2/2 - a_S v_0^2/4
\end{array}
\right).
\end{equation}
The physical Higgses are defined by
\begin{eqnarray}
\label{eq:neutralmix} \left( \begin{array}{c} h \\ H\end{array}
\right) & = & \left(\begin{array}{ccc} \cos \theta_0 & \sin \theta_0
\\ - \sin \theta_0 & \cos \theta_0\end{array} \right) \left(
\begin{array}{c} H^0 \\ S^0 \end{array} \right),
\end{eqnarray}
where the mixing angle is
\begin{equation}
\tan 2 \theta_0 = \frac{a_S v_0 v_S}{ \lambda_H v_0^2 - \lambda_S v_S^2 + a_S (v_0^2 - v_S^2)}.
\end{equation}
It is easy to check that $S_I$ is the Goldstone boson eaten by the
$Z'$ in the theory.
\subsection{Feynman Rules}
We now summarize the Feynman rules for the SM with $U(1)_{B-L}$ and
$U(1)_X$ extensions in Tables~\ref{int} and \ref{int2}, respectively.

\begin{table}[tb]
\begin{center}
\begin{tabular}[t]{|c|c|c|c|}
  \hline
  {\rm Fields} & {\rm Vertices} & \rm Couplings & \rm Approximations\\
  \hline
  $Z'$ & $\bar{q}_iq_iZ'$ & $-iQ^q_{BL}g_{BL}\gamma^\mu$ &$-$\\
       & $q_1=u, q_2=d$   & $Q^q_{BL}={1\over 3}$ &\\
  \hline
       & $\bar{\ell}\ell Z'$   & $-iQ^\ell_{BL}g_{BL}\gamma^\mu$ &$-$\\
       & $\ell=e,\mu,\tau$ & $Q^\ell_{BL}=-1$ &\\
  \hline
       & $\overline{N_{m_1}}N_{m_2}Z'$ & $-i(U_C^TU_C^\ast-V^TV^\ast)_{m_1m_2} Q^\ell_{BL} g_{BL}\gamma^\mu {\gamma_5\over 2}$ & $iI_{m_1m_2} g_{BL}\gamma^\mu {\gamma_5\over 2}$\\
  \hline
       & $\overline{\nu_{m_1}}\nu_{m_2}Z'$ & $-i(U^\dagger U-V_C^\dagger V_C)_{m_1m_2} Q^\ell_{BL} g_{BL}\gamma^\mu {-\gamma_5\over 2}$ & $iI_{m_1m_2} g_{BL}\gamma^\mu {-\gamma_5\over 2}$\\
  \hline
  $N_{m}$ & $\overline{N_m^c}\ell^-W^+$ & $-i{g\over \sqrt{2}}V^\ast_{\ell m}\gamma^\mu P_L$ &$-$\\
          & $N^T_m\ell^-W^+$ & $-i{g\over \sqrt{2}}V^\ast_{\ell m}C\gamma^\mu P_L$ &$-$\\
  \hline
          & $\overline{\nu_{m_1}}N^c_{m_2'}Z$ & $-i{g\over 2c_W}U^{\nu N}_{m_1m_2'}\gamma^\mu P_L$ &$-$\\
          & $\overline{\nu_{m_1}}\overline{N_{m_2'}}^TZ$ & $-i{g\over 2c_W}U^{\nu N}_{m_1m_2'}\gamma^\mu P_LC$ &$-$\\
  \hline
          & $\overline{\nu_\ell}N_mh$ & $-iV_{\ell m}P_R\left({M^N_m\over v_0}c_{\theta_0}+{M^N_m\over v_S}s_{\theta_0}\right)$ & $-iV_{\ell m}P_R{M^N_m\over v_0}c_{\theta_0}$\\
  \hline
            & $\overline{\nu_\ell}N_mH$ & $-iV_{\ell m}P_R\left({M^N_m\over v_0}s_{\theta_0}-{M^N_m\over v_S}c_{\theta_0}\right)$ & $-iV_{\ell m}P_R{M^N_m\over v_0}s_{\theta_0}$\\
  \hline
\end{tabular}
\end{center}
\caption{Feynman rules for $Z'$ and heavy Majorana neutrino $N$ in
SM with $U(1)_{B-L}$ extension, where $U^{\nu N}=U^\dagger V$.}
\label{int}
\end{table}

\subsection{$Z'$ Decays in $U(1)_X$ Extension}

The charge of $U(1)_X$ is defined as $X=Y-5(B-L)/4$ and due to the
mixing between $U(1)_Y$ and $U(1)_X$ we have the mixing matrix of
neutral gauge bosons as below
\begin{eqnarray}
\left(
  \begin{array}{c}
    B^\mu \\
    W_3^\mu \\
    B'^\mu \\
  \end{array}
\right)
=
\left(
  \begin{array}{ccc}
    c_W & -s_Wc' & s_Ws' \\
    s_W & c_Wc' & -c_Ws' \\
    0 & s' & c' \\
  \end{array}
\right) \left(
  \begin{array}{c}
    A^\mu \\
    Z^\mu \\
    Z'^\mu \\
  \end{array}
\right)
\end{eqnarray}
where $s_W(c_W)=\sin\theta_W(\cos\theta_W)$,
$s'(c')=\sin(\theta')(\cos(\theta'))$, and
$\tan(2\theta')=2g_1'\sqrt{g_2^2+g_1^2}/(g_1'^2+25g_{BL}^2{v_S^2\over
v_0^2}-g_2^2-g_1^2)$. $g_1'$ is a free gauge coupling to qualify the
mixing between the two $U(1)$ gauge symmetries. One can get their
mass eigenvalues as
\begin{eqnarray}
M_A&=&0,\\
M_{Z,Z'}&=&{v_0\over 2}\sqrt{g_1^2+g_2^2}\left[{1\over
2}\left({g_1'^2+25({v_S\over v_0})^2g_{BL}^2\over
g_1^2+g_2^2}+1\right)\mp {g_1'\over \sin2\theta'
\sqrt{g_1^2+g_2^2}}\right]^{1/2}.
\end{eqnarray}

The expressions for the possible decays of the $Z'$ are given by
\begin{eqnarray}
\Gamma(Z'\to f\bar{f})&=&{M_{Z'}\over
12\pi}C_f\left[V_f^2\left(1+{2m_f^2\over
M_{Z'}^2}\right)+A_f^2\left(1-{4m_f^2\over
M_{Z'}^2}\right)\right]\sqrt{1-{4m_f^2\over M_{Z'}^2}}
\\
\Gamma(Z'\to \sum_m\nu_m \nu_m)&=&3{M_{Z'}\over
24\pi}C_\nu(X_\nu^U)^2
\\
\Gamma(Z'\to N_m N_m)&=&{M_{Z'}\over
24\pi}C_N(X_N^U)^2\left[1-4{m_N^2\over
M_{Z'}^2}\right]\sqrt{1-{4m_N^2\over M_{Z'}^2}}\\
\Gamma(Z'\to W^+ W^-)&=&{\alpha M_{Z'}s'^2\over
3\tan_W^2}{M_{Z'}^4\over 16M_W^4}\left(1+{16M_W^2\over
M_{Z'}^2}-{68M_W^4\over M_{Z'}^4}-{48M_W^6\over
M_{Z'}^6}\right)\sqrt{1-{4M_W^2\over M_{Z'}^2}}\\
\Gamma(Z'\to Zh)&=&{1\over 48\pi}{M_{Z'}\over
M_Z^2}\left[v_0{c_{\theta_0}\over
4}K-v_Ss_{\theta_0}K'\right]^2\left[1+2{5M_Z^2-M_h^2\over
M_{Z'}^2}+{(M_Z^2-M_h^2)^2\over M_{Z'}^4}\right]\nonumber
\\
&\times&\sqrt{1-2{M_Z^2+M_h^2\over M_{Z'}^2}+{(M_Z^2-M_h^2)^2\over
M_{Z'}^4}}\\
\Gamma(Z'\to ZH)&=&{1\over 48\pi}{M_{Z'}\over
M_Z^2}\left[v_0{s_{\theta_0}\over
4}K+v_Sc_{\theta_0}K'\right]^2\left[1+2{5M_Z^2-M_H^2\over
M_{Z'}^2}+{(M_Z^2-M_H^2)^2\over M_{Z'}^4}\right]\nonumber
\\
&\times&\sqrt{1-2{M_Z^2+M_H^2\over M_{Z'}^2}+{(M_Z^2-M_H^2)^2\over
M_{Z'}^4}}
\end{eqnarray}
where $f=\ell,q$, $C_{\ell,\nu,N}=1,C_q=3$ and
$V_f=(X_{f_L}+X_{f_R})/2,A_f=(-X_{f_L}+X_{f_R})/2$.

\begin{table}[tb]
\begin{center}
\begin{tabular}[t]{|c|c|c|c|}
  \hline
  {\rm Fields} & {\rm Vertices} & \rm Couplings & \rm Approximations\\
  \hline
  $Z'$ & $\bar{u}_Lu_LZ'$ & $iX_{u_L}\gamma^\mu P_L$ &$-$\\
       &&$X_{u_L}=(1-{2\over 3}s_W^2)s'\sqrt{g_2^2+g_1^2}+{1\over 3}c'g_1'+{5\over 4}Q^q_{BL}c'g_{BL}$& \\
        & $\bar{u}_Ru_R Z'$ & $iX_{u_R}\gamma^\mu P_R$ &$-$\\
        &&$X_{u_R}=-{2\over 3}s_W^2s'\sqrt{g_2^2+g_1^2}-{2\over 3}c'g_1'+{5\over 4}Q^q_{BL}c'g_{BL}$& \\
        & $\bar{d}_Ld_L Z'$ & $iX_{d_L}\gamma^\mu P_L$ &$-$\\
        &&$X_{d_L}=(-{1\over 2}+{1\over 3}s_W^2)s'\sqrt{g_2^2+g_1^2}-{1\over 6}c'g_1'+{5\over 4}Q^q_{BL}c'g_{BL}$& \\
        & $\bar{d}_Rd_R Z'$ & $iX_{d_R}\gamma^\mu P_R$ &$-$\\
        &&$X_{d_R}={1\over 3}s_W^2s'\sqrt{g_2^2+g_1^2}+{1\over 3}c'g_1'+{5\over 4}Q^q_{BL}c'g_{BL}$& \\
  \hline
       & $\bar{\ell}_L\ell_L Z'$   & $iX_{\ell_L}\gamma^\mu P_L$ &$-$\\
       &&$X_{\ell_L}=(-{1\over 2}+s_W^2)s'\sqrt{g_2^2+g_1^2}+{1\over 2}c'g_1'+{5\over 4}Q^\ell_{BL}c'g_{BL}$& \\
       & $\bar{\ell}_R\ell_R Z'$   & $iX_{\ell_R}\gamma^\mu P_R$ &$-$\\
       & $\ell=e,\mu,\tau$ & $X_{\ell_R}=s_W^2s'\sqrt{g_2^2+g_1^2}+c'g_1'+{5\over 4}Q^\ell_{BL}c'g_{BL}$ &\\
  \hline
       & $\overline{N_{m_1}}N_{m_2}Z'$ & $i(X_N^UU_C^TU_C^\ast+X_N^VV^TV^\ast)_{m_1m_2}\gamma^\mu {\gamma_5\over 2}$ & $iX_N^UI_{m_1m_2}\gamma^\mu {\gamma_5\over 2}$\\
  \hline
       & $\overline{\nu_{m_1}}\nu_{m_2}Z'$ & $i(X_\nu^UU^\dagger U+X^V_\nu V_C^\dagger V_C)_{m_1m_2}\gamma^\mu {-\gamma_5\over 2}$ & $iX_\nu^UI_{m_1m_2}\gamma^\mu {-\gamma_5\over 2}$\\
       &&$X_N^U=-X^V_\nu={5\over 4}Q^\ell_{BL}c'g_{BL}$&\\
       &&$X_\nu^U=-X^V_N={1\over 2}s'\sqrt{g_2^2+g_1^2}+{1\over 2}c'g_1'+{5\over 4}Q^\ell_{BL}c'g_{BL}$&\\
  \hline
  & $W^-_\mu(p_1)W^+_\nu(p_2)Z'_\rho(p_3)$ & $-igc_Ws'[(p_1-p_2)_\rho g_{\mu\nu}+(p_2-p_3)_\mu g_{\nu\rho}+(p_3-p_1)_\nu g_{\rho\mu}]$ & $-$\\
  \hline
  &$hZ_\mu Z'_\nu$&$2i[v_0{c_{\theta_0}\over 4}K-v_Ss_{\theta_0}K']g_{\mu\nu}$& $-$\\
  &$HZ_\mu Z'_\nu$&$2i[v_0{s_{\theta_0}\over 4}K+v_Sc_{\theta_0}K']g_{\mu\nu}$& $-$\\
  &&$K=-\sin(2\theta')(g_1^2+g_2^2+g_1'^2)-2\cos(2\theta')g_1'\sqrt{g_1^2+g_2^2}$&\\
  &&$K'={25\over 4}\sin(2\theta')g_{BL}^2$&\\
  \hline
\end{tabular}
\end{center}
\caption{Feynman rules for $Z'$ in SM with $U(1)_{X}$ extension,
where $s'(c')=\sin(\theta')(\cos(\theta'))$,
$\tan(2\theta')=2g_1'\sqrt{g_2^2+g_1^2}/(g_1'^2+25g_{BL}^2{v_S^2\over
v_0^2}-g_2^2-g_1^2)$ and all momenta are incoming.} \label{int2}
\end{table}

\end{document}